\documentclass{article}
\usepackage{arxiv}
\usepackage[utf8]{inputenc} 
\usepackage[T1]{fontenc}    
\usepackage[colorlinks,urlcolor=blue,citecolor=blue]{hyperref}
\usepackage{url}            
\usepackage{booktabs}       
\usepackage{amsfonts}       
\usepackage{nicefrac}       
\usepackage{microtype}      
\usepackage{lipsum}		
\usepackage{graphicx}
\usepackage{subcaption}
\usepackage{float}
\usepackage{doi}
\usepackage[style=numeric]{biblatex}  
\usepackage{amsmath}
\usepackage{mathtools}
\usepackage{xcolor}
\usepackage{comment}

\title{EVODMs: variational learning of PDEs for stochastic systems via diffusion models with quantified epistemic uncertainty}

\author{
 Zequn He\thanks{Corresponding authors: \texttt{hezequn@seas.upenn.edu (Zequn He)}, \texttt{creina@seas.upenn.edu (Celia Reina)}} \\
 Department of Mechanical Engineering and Applied Mechanics\\
 University of Pennsylvania\\
 Philadelphia, PA 19104, USA \\
 \And
 Celia Reina\footnotemark[1] \\
 Department of Mechanical Engineering and Applied Mechanics\\
 University of Pennsylvania\\
 Philadelphia, PA 19104, USA \\
}

\date{}

\begin{document}
\maketitle
\begin{abstract}
We present Epistemic Variational Onsager Diffusion Models (EVODMs), a machine learning framework that integrates Onsager's variational principle with diffusion models to enable thermodynamically consistent learning of free energy and dissipation potentials (and associated evolution equations) from noisy, stochastic data in a robust manner. By further combining the model with Epinets, EVODMs quantify epistemic uncertainty with minimal computational cost. The framework is validated through two examples: (1) the phase transformation of a coiled-coil protein, modeled via a stochastic partial differential equation, and (2) a lattice particle process (the symmetric simple exclusion process) modeled via Kinetic Monte Carlo simulations. In both examples, we aim to discover the thermodynamic potentials that govern their dynamics in the deterministic continuum limit. 
EVODMs demonstrate a superior accuracy in recovering free energy and dissipation potentials from noisy data, as compared to traditional machine learning frameworks. Meanwhile, the epistemic uncertainty is quantified efficiently via Epinets and knowledge distillation. This work highlights EVODMs' potential for advancing data-driven modeling of non-equilibrium phenomena and uncertainty quantification for stochastic systems.
\end{abstract}


\section{Introduction}
\label{sec: intro}
The development of data-driven models that integrate physical laws into their learning framework has seen substantial growth driven by advances in scientific machine learning (SciML) \cite{karniadakis2021physics}. In particular, physics-informed neural networks (PINNs) have emerged as a pivotal approach where physical laws are directly embedded in the training process of machine learning models \cite{karniadakis2021physics, raissi2019physics}. Such an approach ensures that the learned models remain consistent with fundamental scientific principles and can achieve in many cases excellent accuracy and strong generalizability while requiring only a small dataset. PINNs have become increasingly crucial in many disciplines such as mechanics \cite{raissi2019physics, cai2021physics, haghighat2021physics}, material science \cite{zhang2022analyses, shukla2021physics}, seismology \cite{karimpouli2020physics, rasht2022physics}, and biology \cite{zhang2024discovering, sahli2020physics}. Despite the success of PINNs, challenges remain. Deep neural networks, as highlighted by \cite{szegedy2013intriguing}, are highly sensitive to imperceptible perturbations in their inputs; even minor noise can lead to entirely incorrect predictions. Consequently, without effective uncertainty quantification (UQ) and demonstrated robustness to handle noisy data, these models cannot be fully trusted. To build trustworthy models, an important aspect is the quantification of epistemic uncertainty, which arises due to incomplete knowledge
\cite{hullermeier2021aleatoric, psaros2023uncertainty}. Established methods for UQ include Gaussian processes (GPs) \cite{rasmussen2003gaussian}, Bayesian methods \cite{neal2012bayesian}, polynomial chaos expansions (PCEs) \cite{najm2009uncertainty}, and ensembles \cite{lakshminarayanan2017simple}. Some of these methods have been integrated with PINNs and show promising results such as Bayesian PINNs (B-PINNs) \cite{yang2021b}, physics-informed PCEs ($\text{PC}^2$) \cite{novak2024physics}, and ensemble PINNs (E-PINNs) \cite{jiang2023practical}. Nonetheless, learning robustly from noisy data generated by stochastic dynamics, as opposed to measurement noise, while also quantifying epistemic uncertainty, remains a significant challenge.

In the past decade, generative models have revolutionized the field of machine learning. They learn the underlying probability distributions of complex datasets, making it possible to generate new samples that closely resemble the training data. Prominent generative models include variational autoencoders (VAEs) \cite{kingma2013auto}, generative adversarial networks (GANs) \cite{goodfellow2014generative}, and normalizing flows (NFs) \cite{rezende2015variational}. These models have been applied across a broad spectrum of tasks, including anomaly detection \cite{an2015variational}, image synthesis \cite{karras2019style}, style transfer \cite{isola2017image}, video generation \cite{vondrick2016generating}, and molecular design \cite{elton2019deep}. Additionally, the SciML community has actively explored embedding physics-based constraints into these generative models for solving partial differential equations (PDEs) and stochastic differential equations (SDEs). For instance, \cite{zhong2023pi, shin2023physics} integrate physics-informed learning with variational inference to address SDEs. Similarly, adversarial inference has been combined with physics-informed constraints to solve PDEs \cite{yang2019adversarial, daw2021pid, gao2022wasserstein} and SDEs \cite{yang2020physics}. Furthermore, physics-informed flow models have been proposed for tackling stochastic forward and inverse problems \cite{guo2022normalizing}. Recently, diffusion models, introduced by \cite{sohl2015deep} and further advanced as denoising diffusion probabilistic models (DDPMs) by \cite{ho2020denoising}, have gained significant attention due to their competitive performance, often surpassing models such as VAEs, GANs, and NFs in a variety of domains. This ranges from computer vision \cite{ho2020denoising, song2020score, dhariwal2021diffusion}, natural language processing \cite{austin2021structured, hoogeboom2021argmax}, and audio analysis \cite{alexanderson2023listen, huang2023make}, to interdisciplinary applications in computational chemistry \cite{hoogeboom2022equivariant, jing2022torsional} and medical image analysis \cite{cao2024high, chung2022score}. In the field of SciML, diffusion models have been used for studies in fluid dynamics \cite{shu2023physics, gao2024bayesian, du2024conditional, bastek2024physics} and have shown promising results. Moreover, prior work on incorporating physical constraints into diffusion models has emerged as a powerful approach for addressing PDEs. For example, CoCoGen \cite{jacobsen2023cocogen} 
enforces physical consistency by embedding the governing equations directly into the sampling process. This approach employs a residual minimization step during the sampling process, which ensures that generated solutions adhere closely to the underlying physics. DiffusionPDE \cite{huang2024diffusionpde} addresses the challenge of solving PDEs under partial observations. Unlike CoCoGen, DiffusionPDE uses a joint generative framework to reconstruct both the coefficients and the solutions of the PDEs simultaneously. This method leverages sparse observations and incorporates PDE-guided denoising during inference, enabling it to solve both forward and inverse problems for static and dynamic PDEs. Similarly, by embedding the physics in the guidance, Pi-fusion \cite{qiu2024pi}, a physics-informed diffusion model, has demonstrated promising results in learning the temporal evolution of velocity and pressure fields in fluid dynamics. On the other hand, physics-informed diffusion models (PIDMs) \cite{bastek2024physics} are proposed to integrate physical laws into diffusion models in a probabilistic manner with the concept of virtual observables that is first proposed in \cite{rixner2021probabilistic}. However, none of these works have demonstrated their applicability to data generated from stochastic processes or their ability to quantify uncertainty. Among the aforementioned studies in SciML, some of them have already demonstrated the effectiveness of generative models like VAEs, GANs, and NFs, in handling data with measurement noise and serving as surrogate models for UQ \cite{shin2023physics, yang2019adversarial, daw2021pid, gao2022wasserstein, guo2022normalizing}.  
Diffusion models have also been recently studied for UQ. In addition to direct conditional sampling \cite{finzi2023user}, efforts have been made to integrate diffusion models with Bayesian inference \cite{lu2024diffusion, fan2024general} and ensemble methods \cite{leinonen2023latent, shu2024zero, asperti2025precipitation}. These approaches highlight the versatility of diffusion models in capturing uncertainty across diverse domains, from dynamical systems to weather prediction. Nonetheless, diffusion models are well-known for being computationally intensive during sampling due to their iterative denoising process. Their combination with Bayesian inference and ensembles increases the computational cost of inference even more, posing a challenge for practical applications of diffusion models for UQ.

Beyond these substantial advancements in SciML, the inclusion of thermodynamic considerations using data-driven approaches is an emerging field of research \cite{teichert2019machine, hernandez2021structure, thakolkaran2022nn, gruber2023reversible, fuhg2024polyconvex, rittig2024thermodynamics}. Of relevance to this work is Variational Onsager Neural Networks (VONNs), which leverage Onsager’s variational principle to learn non-equilibrium evolution equations while strongly encoding thermodynamic consistency \cite{huang2022variational}. This approach focuses on learning the free energy and dissipation potential densities from spatial-temporal data of macroscopic observables, while the dynamics directly follow from these and the variational statement. Although VONNs have proven to be a successful learning strategy, their accuracy can be hindered in the presence of noisy data \cite{qiu2025bridging}. Meanwhile, since deterministic neural networks are used as backbones for VONNs, they naturally lack the capability to quantify uncertainty. As a result, the applicability and reliability of VONNs are significantly constrained in scenarios dominated by stochastic dynamics, where data is inherently noisy. 

In this study, to address the existing limitations of VONNs, we propose a framework named Epistemic Variational Onsager Diffusion Models (EVODMs). EVODMs are designed to learn the free energy and dissipation potentials from noisy data with the second law of thermodynamics strongly enforced. The governing equations derived from Onsager's variational principle are used as the physics constraints in the loss functions for training. Additionally, instead of using Bayesian inference and ensembles, we integrate a novel UQ framework called Epinet \cite{osband2023epistemic} into our model for efficient and reliable quantification of epistemic uncertainty, adding only minimal computational overhead. Finally, we train the Epinets component of EVODMs through knowledge distillation \cite{hinton2015distilling} to further accelerate and stabilize the training process. 

The effectiveness of EVODMs is validated through two examples: the phase transformation of a coiled-coil protein and the symmetric simple exclusion process (SSEP). In the first example, EVODMs accurately learn the non-convex free energy and the quadratic dissipation potential of a protein undergoing phase transformation modeled by overdamped Langevin dynamics. Here, the noise physically originates from the interaction of the protein with the solvent. The second example studied, SSEP, is a lattice process that is evolved in time via Kinetic Monte Carlo simulations. Despite the inherent lack of uniqueness in the free energy and dissipation potentials, EVODMs successfully discover the system's dynamics in the continuum limit from noisy data generated by the particle simulations. In both examples, the model performances are compared to VONNs, which reveals the enhanced capability of EVODMs in handling noisy data. Moreover, EVODMs provide estimates of epistemic uncertainty, which confirm their robustness and reliability in addressing complex, stochastic systems.

The paper is organized as follows: Section~\ref{sec: preliminary} provides an overview of DDPMs, VONNs, and Epistemic Neural Networks (ENNs). The proposed EVODMs are introduced in detail in Section~\ref{sec: evodm}. Section~\ref{sec: experiments} presents the two examples studied, including the model descriptions, data generation, data preprocessing, neural network training and loss functions, as well as the corresponding results. Finally, conclusions and future directions are outlined in Section~\ref{sec: conclusion}.

\section{Preliminary}
\label{sec: preliminary}
\subsection{Denoising Diffusion Probabilistic Models (DDPMs)}
\label{sec: ddpm}
In DDPMs, the model progressively transforms a sample from a simple prior, typically a standard Gaussian distribution, into a sample from an unknown data distribution $q(\mathbf{y})$ \cite{ho2020denoising}. In the forward diffusion process, we start with real data $\mathbf{y}_0 \sim q\left(\mathbf{y}_0\right)$ and sequentially add small amounts of Gaussian noise over $S$ timesteps. This process is defined by the Markov chain,
\begin{equation}
q\left(\mathbf{y}_{1: S} \mid \mathbf{y}_0\right)=\prod_{s=1}^S q\left(\mathbf{y}_s \mid \mathbf{y}_{s-1}\right),
\end{equation}
where each transition adds noise according to
\begin{equation}
q\left(\mathbf{y}_s \mid \mathbf{y}_{s-1}\right)=\mathcal{N}\left(\mathbf{y}_s ; \sqrt{1-\beta_s} \mathbf{y}_{s-1}, \beta_s \mathbf{I}\right),
\end{equation}
and $\beta_s \in(0,1)$ is a variance schedule controlling the amount of noise at each step. Due to the properties of Gaussian distributions, we can express the distribution of $\mathbf{y}_s$ conditioned directly on $\mathbf{y}_0$ as
\begin{equation}
q\left(\mathbf{y}_s \mid \mathbf{y}_0\right)=\mathcal{N}\left(\mathbf{y}_s ; \sqrt{\bar{\alpha}_s} \mathbf{y}_0,\left(1-\bar{\alpha}_s\right) \mathbf{I}\right),
\end{equation}
where $\alpha_s=1-\beta_s$ and $\bar{\alpha}_s=\prod_{l=1}^s \alpha_l$. As $s$ approaches $S$,  $\bar{\alpha}_S$ becomes very small and $\mathbf{y}_S$ approximates an isotropic Gaussian distribution.

In the reverse diffusion process, we define a parameterized Markov chain that aims to invert the forward process, namely,
\begin{equation}
q\left(\mathbf{y}_{0: S}\right)=p\left(\mathbf{y}_S\right) \prod_{s=1}^S q\left(\mathbf{y}_{s-1} \mid \mathbf{y}_s\right),
\end{equation}
where $p\left(\mathbf{y}_S\right)$ is $\mathcal{N}\left(\mathbf{0}, \mathbf{I}\right)$ and $q\left(\mathbf{y}_{s-1} \mid \mathbf{y}_s\right)$ is approximated by a neural network such that
\begin{equation}
p_{\boldsymbol{\theta}}\left(\mathbf{y}_{s-1} \mid \mathbf{y}_s\right)=\mathcal{N}\left(\mathbf{y}_{s-1} ; \boldsymbol{\mu_\theta}\left(\mathbf{y}_s, s\right), \sigma^2_q(s) \mathbf{I}\right).
\end{equation}
Here, $\boldsymbol{\mu_\theta}\left(\mathbf{y}_s, s\right)$ is a neural network with parameters $\boldsymbol{\theta}$ that predicts the mean of the reverse distribution, and ${\sigma_q}^2 (s)$ is typically set to match the variance of the forward process and is given by the choice of noise scheduler. Sometimes, it is treated as a learned parameter as mentioned in \cite{nichol2021improved}.

The training objective of DDPMs involves minimizing the variational bound of the negative log-likelihood of the data. This bound $\mathcal{L}_{\boldsymbol{\theta}}$ decomposes into a sum of Kullback–Leibler (KL) divergences and a reconstruction term \cite{ho2020denoising}, 
\begin{equation}
\label{eq: elbo}
\mathcal{L}_{\boldsymbol{\theta}}=\mathbb{E}_q\left[\underbrace{D_{\mathrm{KL}}\left(q\left(\mathbf{y}_S \mid \mathbf{y}_0\right) \| p\left(\mathbf{y}_S\right)\right)}_{\mathcal{L}_S}+\sum_{s=2}^S \underbrace{D_{\mathrm{KL}}\left(q\left(\mathbf{y}_{s-1} \mid \mathbf{y}_s, \mathbf{y}_0\right) \| p_{\boldsymbol{\theta}}\left(\mathbf{y}_{s-1} \mid \mathbf{y}_s\right)\right)}_{\mathcal{L}_{s-1}}-\underbrace{\ln p_{\boldsymbol{\theta}}\left(\mathbf{y}_0 \mid \mathbf{y}_1\right)}_{\mathcal{L}_0}\right],
\end{equation}
where $\mathcal{L}_S$ serves as the prior matching term that measures how closely the distribution of the noised data at the step $S$, $q\left(\mathbf{y}_S \mid \mathbf{y}_0\right)$, is to the assumed prior distribution $p\left(\mathbf{y}_S\right)$, typically modeled as an isotropic Gaussian. $\mathcal{L}_{s-1}$ are the reverse KL terms that encourage the learned reverse transitions to approximate the true posterior transitions of the forward process. Finally, $\mathcal{L}_0$ is the reconstruction term that evaluates how well the model can reconstruct the original data from the first noised version $\mathbf{y}_1$.

The distribution $q\left(\mathbf{y}_{s-1} \mid \mathbf{y}_s, \mathbf{y}_0\right)$ in Equation~\eqref{eq: elbo} can be written as
\begin{equation}
q\left(\mathbf{y}_{s-1} \mid \mathbf{y}_s, \mathbf{y}_0\right)=\mathcal{N}\left(\mathbf{y}_{s-1} ; \boldsymbol{\mu}_q\left(\mathbf{y}_s, \mathbf{y}_0\right), \sigma^2_q(s) \mathbf{I}\right),
\end{equation}
where
\begin{equation}
\boldsymbol{\mu}_q\left(\mathbf{y}_s, \mathbf{y}_0\right)=\frac{\left(1-\bar{\alpha}_{s-1}\right) \sqrt{\alpha_s}}{1-\bar{\alpha}_s} \mathbf{y}_s+\frac{\left(1-\alpha_s\right) \sqrt{\bar{\alpha}_{s-1}}}{1-\bar{\alpha}_s} \mathbf{y}_0,
\end{equation}
and
\begin{equation}
\sigma_q^2(s)=\frac{\left(1-\alpha_s\right)(1-\sqrt{\bar{\alpha}_{s-1}})}{1-\bar{\alpha}_s}.
\end{equation}
Here, $q\left(\mathbf{y}_{s-1} \mid \mathbf{y}_s, \mathbf{y}_0\right)$ is completely determined by $\mathbf{y}_s$ and $\mathbf{y}_0$, hence, no neural network is required to estimate the mean and variance. The reverse KL divergence in Equation~\eqref{eq: elbo} is then simplified to
\begin{equation}
D_{\mathrm{KL}}\left(q\left(\mathbf{y}_{s-1} \mid \mathbf{y}_s, \mathbf{y}_0\right) \| p_{\boldsymbol{\theta}}\left(\mathbf{y}_{s-1} \mid \mathbf{y}_s\right)\right) = \frac{1}{2 \sigma_q^2(s)}\left\|\boldsymbol{\mu}_q\left(\mathbf{y}_s, \mathbf{y}_0\right)-\boldsymbol{\mu}_{\boldsymbol{\theta}}\left(\mathbf{y}_s\right)\right\|^2.
\end{equation}
In particular, $\boldsymbol{\mu}_{\boldsymbol{\theta}}$ can be defined as
\begin{equation}
\boldsymbol{\mu}_{\boldsymbol{\theta}}\left(\mathbf{y}_s\right) = \frac{\left(1-\bar{\alpha}_{s-1}\right) \sqrt{\alpha_s}}{1-\bar{\alpha}_s} \mathbf{y}_s+\frac{\left(1-\alpha_s\right) \sqrt{\bar{\alpha}_{s-1}}}{1-\bar{\alpha}_s} \hat{\mathbf{y}}_{\boldsymbol{\theta}}\left(\mathbf{x}_t\right).
\end{equation}
Therefore, Equation (\ref{eq: elbo}) (omitting $\mathcal{L}_S$ as it is independent of $\boldsymbol{\theta}$), can be simplified to
\begin{equation}
\mathcal{L}_{\boldsymbol{\theta}}=\sum_{s=1}^S \mathbb{E}_{q\left(\mathbf{y}_s \mid \mathbf{y}_0\right)}\left[\frac{1}{2 \sigma_q^2(s)} \frac{\left(1-\alpha_s\right)^2 \bar{\alpha}_{s-1}}{\left(1-\bar{\alpha}_s\right)^2}\left\|\hat{\mathbf{y}}_{\boldsymbol{\theta}}\left(\mathbf{y}_s, s\right)-\mathbf{y}_0\right\|^2\right],
\end{equation}
where $\hat{\mathbf{y}}_{\boldsymbol{\theta}}$ is the denoising network that is trained for all noisy conditions. More detailed derivations of DDPMs can be found in \cite{chan2024tutorial}.

For the purpose of this study, we here introduce conditional DDPMs aiming to model the distribution $p_{\boldsymbol{\theta}}(\mathbf{y}_0 \mid \mathbf{x})$, where $\mathbf{x} \in \mathcal{X}$ is some condition or auxiliary information. The forward process $q\left(\mathbf{y}_{1: S} \mid \mathbf{y}_0\right)$ remains the same as in the unconditional case. However, conditioning modifies the reverse (denoising) distribution $p_{\boldsymbol{\theta}}\left(\mathbf{y}_{s-1} \mid \mathbf{y}_s, \mathbf{x}\right)$, making it depend not only on the noisy sample $\mathbf{y}_s$ but also on $\mathbf{x}$. The reverse distribution now includes $\mathbf{x}$ and approximated by the neural network turns into
\begin{equation}
p_{\boldsymbol{\theta}}\left(\mathbf{y}_{0: S} \mid \mathbf{x}\right)=p\left(\mathbf{y}_S\right) \prod_{s=1}^S p_{\boldsymbol{\theta}}\left(\mathbf{y}_{s-1} \mid \mathbf{y}_s, \mathbf{x}\right),    
\end{equation}
where
\begin{equation}
p_{\boldsymbol{\theta}}\left(\mathbf{y}_{s-1} \mid \mathbf{y}_s, \mathbf{x}\right)=\mathcal{N}\left(\mathbf{y}_{s-1} \mid \boldsymbol{\mu_\theta}\left(\mathbf{y}_s, s, \mathbf{x}\right), \sigma^2_q(s) \mathbf{I}\right).
\end{equation}
Here, the mean $\boldsymbol{\mu_\theta}(\cdot)$ is computed by a neural network conditioned on $\mathbf{y}_s$, the timestep $s$, and the condition $\mathbf{x}$. 
Analogously to the unconditional case, the loss function used for training a conditional DDPM to obtain the optimal model parameters $\boldsymbol{\theta}^*$ is thus
\begin{equation}
\label{eq: train loss DDPM}
\boldsymbol{\theta}^*=\underset{\boldsymbol{\theta}}{\operatorname{argmin}} \sum_{s=1}^S \frac{1}{2 \sigma_q^2(s)} \frac{\left(1-\alpha_s\right)^2 \bar{\alpha}_{s-1}}{\left(1-\bar{\alpha}_s\right)^2} \mathbb{E}_{q\left(\mathbf{y}_s \mid \mathbf{y}_0\right)}\left[\left\|\hat{\mathbf{y}}_\theta\left(\mathbf{y}_s,s,\mathbf{x}\right)-\mathbf{y}_0\right\|^2\right].
\end{equation}
In practice, the summation $\sum_{s=1}^S$ is replaced by a uniform distribution $s \sim$ Uniform $[1, S]$. Furthermore, the expectation $\mathbb{E}_{q\left(\mathbf{y}_s \mid \mathbf{y}_0\right)}$ is approximated by Monte Carlo samples (typically implemented as a mini-batch) from $q\left(\mathbf{y}_s \mid \mathbf{y}_0\right)$, and therefore, the explicit expectation from the training objective is removed. The term $\frac{1}{2 \sigma_q^2(s)} \frac{\left(1-\alpha_s\right)^2 \bar{\alpha}_{s-1}}{\left(1-\bar{\alpha}_s\right)^2}$ is controlled by the noise scheduler. For simplicity, we can drop this term as its impact is minor \cite{chan2024tutorial}. Consequently, Equation~\eqref{eq: train loss DDPM} simplifies to
\begin{equation}
\label{eq: single timestep train loss DDPM}
\boldsymbol{\theta}^*=\underset{\boldsymbol{\theta}}{\operatorname{argmin}} \left\|\hat{\mathbf{y}}_{\boldsymbol{\theta}}\left(\mathbf{y}_s,s,\mathbf{x}\right)-\mathbf{y}_0\right\|^2,
\end{equation}
which is the training loss used at a single time step $s$ that is drawn from a uniform distribution over $[1,S]$. Each training iteration optimizes this objective for Monte Carlo samples $\left\{\left(\mathbf{y}_s, s, \mathbf{x}, \mathbf{y}_0\right)\right\}$. Over many iterations, this approximates the loss function. However, in this work, the neural network outputs do not directly represent the target quantities and, as a result, Equation~\eqref{eq: single timestep train loss DDPM} is not used during training. The proposed EVODM is trained by imposing the physical laws in a similar manner. The detailed formulation of these loss functions will be introduced in Section~\ref{sec: PDE constraints cond}.

\subsection{Variational Onsager Neural Networks (VONNs)}
\label{sec: vonns}
Consider an isothermal system where inertia is negligible. The Rayleighian, $\mathcal{R}$, is a functional that characterizes the system's dynamics. It is expressed as
\begin{equation}
\mathcal{R}[\mathbf{z}, \mathbf{w}]=\dot{\mathcal{F}}[\mathbf{z}, \mathbf{w}]+\mathcal{D}[\mathbf{z}, \mathbf{w}]+\mathcal{P}[\mathbf{z}, \mathbf{w}],
\end{equation}
where $\mathbf{z}$ represents the state variables, and $\mathbf{w}$ denotes the process variables (these are related to $\dot{\mathbf{z}}$ and more directly describe how the system dissipates energy). The Rayleighian is composed of $\dot{\mathcal{F}}[\mathbf{z}, \mathbf{w}]$, which is the time derivative of the system's free energy, $\mathcal{D}[\mathbf{z}, \mathbf{w}]$, the dissipation potential, and $\mathcal{P}[\mathbf{z}, \mathbf{w}]$ the power supplied by the external forces. The dissipation potential is required to satisfy (i) $\mathcal{D}[\mathbf{z}, \mathbf{w}]$ convex with respect to $\mathbf{w}$, (ii) $\mathcal{D}[\mathbf{z}, \mathbf{0}]=0$, and (iii) $\mathcal{D}$ reaches its minimum value at $\mathbf{w}=0$, which ensures compliance with the second law of thermodynamics \cite{mielke2016generalization, kraaij2020fluctuation, arroyo2018onsager}. According to Onsager's variational principle \cite{arroyo2018onsager, doi2011onsager}, the evolution of such a system is determined by minimizing the Rayleighian with respect to $\mathbf{w}$, i.e., 
\begin{equation}
    \min _{\mathbf{w}} \mathcal{R}[\mathbf{z}, \mathbf{w}],
\end{equation}
which yields the following governing equations
\begin{equation} 
\label{eq: thermo}
    \frac{\delta \dot{\mathcal{F}}}{\delta \mathbf{w}} +\frac{\delta \mathcal{D}}{\delta \mathbf{w}}+\frac{\delta \mathcal{P}}{\delta \mathbf{w}}=\mathbf{0}.
\end{equation}

The described formalism has been successfully integrated with neural networks in \cite{huang2022variational}, forming a thermodynamics-based variational learning strategy for non-equilibrium PDEs known as VONNs. Two assumptions are made in the original formulation of VONNs, which we also consider in this study. First, the state and process variables are assumed to be known and measurable (recent extensions have been made to partially relax this assumption \cite{qiu2025bridging}). Second, it is assumed, as is common to many continuum models, that both the free energy and dissipation potential possess associated densities. Specifically, the free energy is represented as $\mathcal{F}[\mathbf{z}]=\int_{\Omega} f(\mathbf{z})\, dV$ , and the dissipation potential is expressed as $\mathcal{D}[\mathbf{z}, \mathbf{w}]=\int_{\Omega} \psi(\mathbf{z}, \mathbf{w})\, dV$, where $\Omega$ is the reference domain. In VONNs, the free energy density $f(\mathbf{z})$ and the dissipation potential density $\psi(\mathbf{z}, \mathbf{w})$ are simultaneously learned using separate neural networks that strictly encode the aforementioned conditions for thermodynamic consistency.

For the purposes of this investigation, we will use the reformulation of VONNs, proposed in \cite{qiu2025bridging}, which models instead the free energy density and dual dissipation potential density. The latter is defined from $\psi$ via Legendre transform as 
\begin{equation}
    \phi(\mathbf{z}, \mathbf{g})
    =\sup _{\mathbf{w}}(\mathbf{w} \cdot \mathbf{g}-\psi(\mathbf{z}, \mathbf{w})),
\end{equation}
where $\mathbf{g}=\partial \psi / \partial \mathbf{w}$ represents the conjugate force corresponding to $\mathbf{w}$. This formulation allows to write the PDEs resulting from Onsager's variational principle, of the form $
    \frac{\partial \psi(\mathbf{z},\mathbf{w})}{\partial \mathbf{w}}
    = \mathcal{H}(f)(\mathbf{z})$, where $\mathcal{H}$ is an operator, as
\begin{equation}
    \mathbf{w} = \left. \frac{\partial \phi(\mathbf{z}, \mathbf{g})}{\partial \mathbf{g}} \right|_{\mathbf{g} = \mathcal{H}(f)(\mathbf{z})}
    = \nabla_{\mathbf{g}} \phi (\mathbf{z}, \mathcal{H}(f)(\mathbf{z})).
\end{equation}
We recall that $\mathbf{w}$ is an observable, making this formulation of VONNs more flexible for integration with DDPMs. The conditions on the dissipation potential density translate into analogous conditions for its dual. That is, $\phi$ is required to satisfy (i) $\phi(\mathbf{z},\mathbf{g})$ convex with respect to $\mathbf{g}$, (ii) $\phi(\mathbf{z}, \mathbf{0}) = 0$, and (iii) $\left.\frac{ \partial \phi }{\partial \mathbf{g}} \right|_{\mathbf{g}=\mathbf{0}} = \mathbf{0}$. Similarly to the original formulation of VONNs, these constraints are strongly enforced in this work by defining the dual dissipation potential density as
\begin{equation}
\label{eq: dualphi}
\phi(\mathbf{z}, \mathbf{g})=\tilde{\phi}(\mathbf{z}, \mathbf{g})-\tilde{\phi}(\mathbf{z}, \mathbf{0})-\left.\frac{\partial \tilde{\phi}}{\partial \mathbf{g}}\right|_{\mathbf{g}=\mathbf{0}} \cdot \mathbf{g},
\end{equation}
where $\tilde{\phi}$ is the output of a Partially Input Convex Integrable Neural Network (PICINN), which integrates the features of an Integrable Neural Network (INN) \cite{teichert2019machine} and those of a Partially Input Convex Neural Network \cite{amos2017input}. 
More specifically, INNs are used as only the partial derivatives of $\phi$ are involved in the training process. The requirement of convexity with respect to $\mathbf{g}$ is met through architectural adaptations based on PICNNs. We remark that even in cases where $\phi$ is only a function of $\mathbf{g}$, we cannot use Fully Input Convex Integrable Neural Networks (FICINNs) as the time embeddings required by the nature of conditional diffusion models will be part of the neural network's input. The dissipation potential density $\psi$ can then be obtained, if so desired, from its dual function $\phi$ as follows
\begin{equation}
\psi(\mathbf{z}, \mathbf{w})=\mathbf{w} \cdot \mathbf{g}-\phi(\mathbf{z}, \mathbf{g}).
\end{equation}
As for the free energy density, this is modeled via an INN. Denoting the output of this INN by $\tilde{f}$, the free energy density is defined as
\begin{equation}
f(\mathbf{z})=\tilde{f}(\mathbf{z})-\tilde{f}(\mathbf{0}).
\end{equation}
Interested readers are referred to Appendix~\ref{appendix: inn free energy density} and Appendix~\ref{appendix: picinn dissipation potential density} for a more detailed explanation of INNs and PICINNs as well as the illustrations of these neural networks. 

\subsection{Epistemic Neural Networks (ENNs)}
\label{sec: enn}
ENNs, introduced in \cite{osband2023epistemic}, provide a versatile framework for generating joint predictions and enable the quantification of epistemic uncertainty. The output of a traditional neural network assigns a corresponding probability to each label. This predictive distribution is known as a marginal prediction, as it pertains to a single input. However, the marginal prediction assesses each outcome independently, which limits its ability to distinguish between epistemic uncertainty and aleatoric uncertainty. To quantify epistemic uncertainty, we need to analyze the dependencies between the outcomes produced by the model and this is where joint prediction is involved and what ENNs are capable of. By making joint predictions, ENNs can efficiently capture and quantify epistemic uncertainty \cite{osband2023epistemic, wen2021predictions, osband2022neural}. 
\begin{figure}[!htbp]
    \centering
    \includegraphics[width=0.5\textwidth]{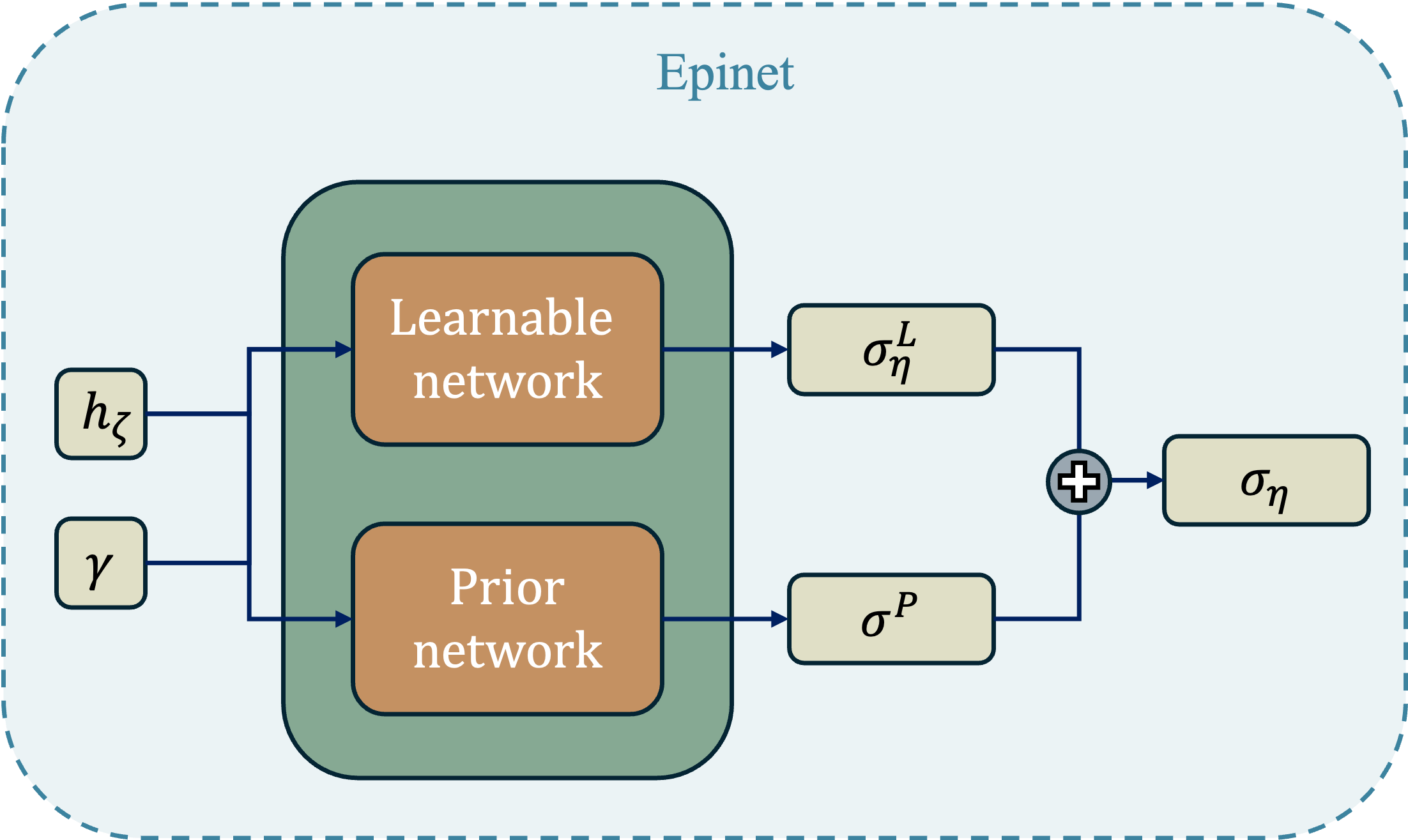}
    \caption{Illustration of an Epinet, which can be added to any base network to convert it into an epistemic neural network. The base network, parameterized by $\zeta$, generates features $h_\zeta(x)$ from the input $x$. These features, which include $x$, and the epistemic indices $\gamma$ are the inputs of the Epinet. The two main components of the Epinet are the learnable network ($\sigma_\eta^L$), parameterized by $\eta$, and the prior network ($\sigma^P$). The output of the Epinet, $\sigma_{\eta}$, combines the predictions of the learnable and prior networks, to quantify epistemic uncertainty.}
    \label{fig: Epinet scheme}
\end{figure}
The core component of an ENN is the Epinet, illustrated in Figure~\ref{fig: Epinet scheme}. The Epinet serves as an extension module that transforms any base neural network into an ENN in a simple and computationally efficient manner 
\begin{equation}
\label{eq: enn equation}
\underbrace{y_\theta(x, \gamma)}_{\text {ENN}}=\underbrace{\mu_\zeta(x)}_{\text {Base network}}+\underbrace{\sigma_\eta\left(\mathrm{sg}\left[h_\zeta(x)\right], \gamma \right)}_{\text {Epinet}},
\end{equation}
where $y_\theta(x, \gamma)$ is the quantity of interest (QoI), $\zeta$ are the parameters of the base network, $\eta$ those of the Epinet, and $\theta=(\zeta,\eta)$ represents the complete set of trainable parameters. The operator $\mathrm{sg}[\cdot]$ represents a "stop gradient" operation, ensuring that the argument remains fixed when computing a gradient. The features $h_\zeta(x)$ are extracted from the base network, often comprising the last hidden layer concatenated with the input $x$. The epistemic index $\gamma$ is sampled from the distribution $P_\Gamma$, with typical choices being a uniform distribution or a standard Gaussian. Since the base network is independent of $\gamma$, the variation in the ENN output is introduced by the Epinet which takes the following form
\begin{equation}
\underbrace{\sigma_\eta(\mathrm{sg}\left[h_\zeta(x)\right], \gamma)}_{\text {Epinet}}=\underbrace{\sigma_\eta^L(\mathrm{sg}\left[h_\zeta(x)\right], \gamma)}_{\text{Learnable network}}+\underbrace{\sigma^P(\mathrm{sg}\left[h_\zeta(x)\right], \gamma)}_{\text{Prior network}},
\end{equation}
where $\sigma_\eta^L$ is the learnable network 
and $\sigma^P$ is the prior network. The prior network is implemented as a fixed ensemble of small neural networks initialized with random weights. These small neural networks then become fixed and not trained. The learnable network $\sigma_\eta^L$ is usually initialized to output values near zero but is later trained to ensure that the resulting sum, $\sigma_\eta$, produces statistically valid predictions for all plausible values of $\gamma$. Variations in a prediction $\sigma_\eta = \sigma_\eta^L + \sigma^P$ at a given input $x$ as a function of $\gamma$ represent the predictive epistemic uncertainty of the model. 

During training, the regression loss for an ENN is similar to that of standard neural networks with the additional average over the epistemic index $\gamma$, i.e., 
\begin{equation}
\label{eq: enn_general_loss}
\mathcal{L}_{\text{ENN}}(\theta)=\frac{1}{|\Gamma|} \sum_{\gamma \in \Gamma}\left(\frac{1}{|\mathcal{D}|} \sum_{(x, y) \in \mathcal{D}} (y_\theta(x, \gamma)-y(x))^2 \right),
\end{equation}
where $|\Gamma|$ is the total number of epistemic indices, $|\mathcal{D}|$ is the number of data points, and $y$ is the ground truth. 

Well-established UQ methods such as deep ensembles \cite{lakshminarayanan2017simple, rahaman2021uncertainty, fort2019deep} and Bayesian neural networks (BNNs) \cite{neal2012bayesian, kononenko1989bayesian, gelman1995bayesian} can also be described as ENNs but not vice versa. A detailed theoretical proof for this argument is provided in \cite{osband2023epistemic}. 

\section{Epistemic Variational Onsager Diffusion Models (EVODMs)}
\label{sec: evodm}
\begin{figure}[!htbp]
    \centering
    \includegraphics[width=0.9\textwidth]{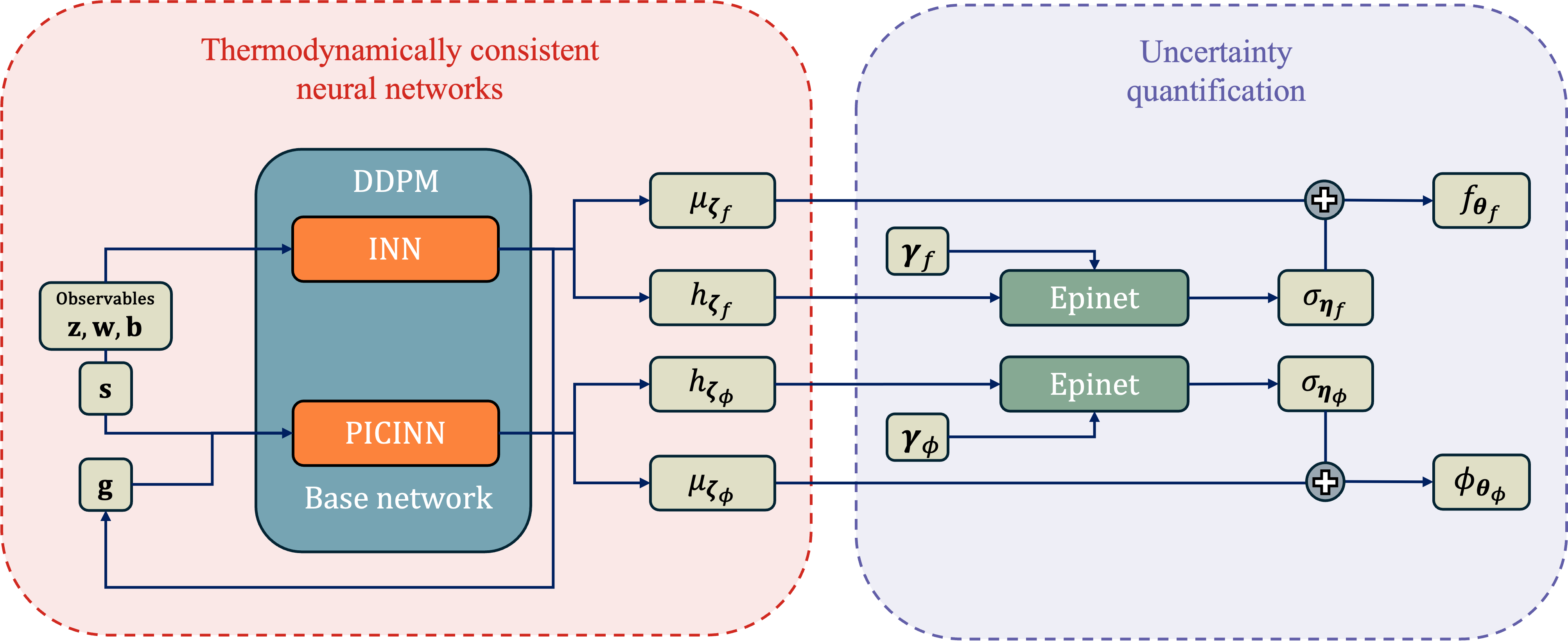}
    \caption{General architecture of an Epistemic Variational Onsager Diffusion Model (EVODM). It combines a Denoising Diffusion Probabilistic Model (DDPM), an Integrable Neural Network (INN), and a Partially Input Convex Integrable Neural Network (PICINN) to model the free energy and dual dissipation potential densities, together with Epinets to quantify the epistemic uncertainty. On the left, the inputs are the time ($\mathbf{s}$) used in DDPM and observables that include the state variables ($\mathbf{z}$), the process variables ($\mathbf{w}$), the boundary data ($\mathbf{b}$) (if applicable), and the conjugate forces ($\mathbf{g}$). The DDPM outputs the free energy and dual dissipation potential densities ($\mu_{\boldsymbol{\zeta}_f}$ and $\mu_{\boldsymbol{\zeta}_\phi}$, respectively), with corresponding features ($h_{\boldsymbol{\zeta}_f}$ and $h_{\boldsymbol{\zeta}_\phi}$) to be used in the Epinets. On the right, $\boldsymbol{\gamma}_f$ and $\boldsymbol{\gamma}_{\phi}$ represent the epistemic indices sampled from their reference distribution (here taken to be the same for both epistemic indices) $P_\Gamma$ and used for the corresponding Epinet. Each Epinet generates uncertainty estimates ($\sigma_{\boldsymbol{\eta}_f}$ and $\sigma_{\boldsymbol{\eta}_\phi}$), which are combined with the base network outputs to produce the learned free energy and dual dissipation potential densities ($f_{\boldsymbol{\theta}_f}$ and $\phi_{\boldsymbol{\theta}_{\phi}}$). The overall set of trainable parameters are thus $\boldsymbol{\theta}_f=(\boldsymbol{\zeta}_f, \boldsymbol{\eta}_f)$ and $\boldsymbol{\theta}_\phi=(\boldsymbol{\zeta}_\phi, \boldsymbol{\eta}_\phi)$, which include the parameters of the base networks $\boldsymbol{\zeta}_f$ and $\boldsymbol{\zeta}_\phi$, and those of the Epinets $\boldsymbol{\eta}_f$ and $\boldsymbol{\eta}_\phi$.}
    \label{fig: evodm scheme}
\end{figure}
The general architecture of the proposed EVODMs is shown in Figure~\ref{fig: evodm scheme}. An EVODM has two primary modules: the first is a conditional DDPM integrated with the thermodynamics-based variational learning strategy detailed in Section~\ref{sec: vonns}. The second module consists of two Epinets, designed to quantify epistemic uncertainty of the learned free energy and dissipation potential densities, with its framework and implementation discussed comprehensively in Section~\ref{sec: epinet}.

In the first module of an EVODM, similarly to VONNs \cite{huang2022variational}, an INN and a PICINN are utilized to ensure integrability of the free energy and dual dissipation potential densities and enforce convexity of the latter with respect to the process variables. For the purpose of function approximation, we use a conditional DDPM, where each neural network directly outputs the corresponding densities and is conditioned on inputs such as $\mathbf{z}$ and $\mathbf{g}$. Although classifier guidance \cite{dhariwal2021diffusion} and classifier-free guidance \cite{ho2022classifier} are widely used conditioning techniques in diffusion models, this work employs explicit conditioning. Specifically, the condition is directly concatenated with the other inputs, which is the most straightforward approach to conditioning in diffusion models.

Unlike most work that uses diffusion models, which typically require the model output as an input for reconstruction, we take a variational approach to learning the evolution equations. That is, while the model outputs the free energy and dual dissipation potential densities which are unobservables, the training of EVODMs relies exclusively on the observable data, namely the state and process variables (and boundary data, if available). Therefore, noise is added to the process variables during the forward process of EVODMs. If boundary data is available, noise is added to them as well. By utilizing automatic differentiation \cite{paszke2017automatic}, the derivatives of the target densities can be computed and integrated into the physics-informed loss functions outlined in Section~\ref{sec: PDE constraints cond}. 
Since the direct outputs of the EVODM are unobservables, no classical data loss term is used and the training framework is instead driven by the physics-informed loss functions.

\subsection{PDE constraints and conditioning}
\label{sec: PDE constraints cond}
Recall that in Section \ref{sec: vonns}, we introduced VONNs and their reformulation, which replaces the learning of dissipation potential density $\psi$ with the learning of its dual function $\phi$ via the Legendre transform. Similar to VONNs, we here embed the governing equations derived from Onsager's variational principle into the loss functions for training the base network of the EVODM in a physics-informed manner. In particular, for the variational learning of $f$ and $\phi$, we consider the following loss function 
\begin{equation}
\label{eq: total loss function}
\mathcal{L}_{\text{Base}}=\lambda_{\text{PDE}} \cdot \mathcal{L}_{\text{PDE}}+\lambda_{\text{BCs}} \cdot \mathcal{L}_{\text{BCs}},
\end{equation}
where $\lambda_{\text{PDE}}$ and $\lambda_{\text{BCs}}$ are the loss weights to control the convergence rate of the corresponding loss terms, associated to the PDE and boundary conditions (if available), respectively. Here, the loss term $\mathcal{L}_{\text{PDE}}$ is defined as
\begin{equation}
\mathcal{L}_{\mathrm{PDE}}=\mathbb{E}_{\substack{(\mathbf{z}, \mathbf{w}) \sim \mathcal{D_{\text{PDE}}}, \\
s \sim  \text{Uniform}[1,S]}}\left[\left\|\mathbf{w}-\frac{\partial \phi\left(\mathbf{w}_{s}, \mathbf{z}, \mathbf{g}(\mathbf{z}, s;\boldsymbol{\zeta}_f),s; \boldsymbol{\zeta}_{\phi}\right)}{\partial \mathbf{g}}\right\|^2\right],
\end{equation}
where $\mathcal{D}_{\text{PDE}}$ is the finite dataset with all the spatial-temporal collocation points for the PDE and $\mathbf{w}_s$ is the noised $\mathbf{w}$ in the forward process at time $s$. The loss term $\mathcal{L}_{\text{BCs}}$, for the case of traction boundary conditions, is
\begin{equation}
\mathcal{L}_{\mathrm{BCs}}=\mathbb{E}_{\substack{\left(\mathbf{z}, \mathbf{b} \right) \sim \mathcal{D}_{\text{BCs}}, \\
s\sim\text{Uniform}[1,S]}}\left[\left\|\mathbf{b}-\frac{\partial f\left(\mathbf{b}_s,\mathbf{z} , s; \boldsymbol{\zeta}_f\right)}{\partial \mathbf{z}}\right\|^2\right],
\end{equation}
here $\mathcal{D}_{\text{BCs}}$ is the dataset with the spatial-temporal data points collected on the boundary. For the training of diffusion models, the inputs of the neural networks also include the noised version of $\mathbf{w}$ and $\mathbf{b}$ from the forward process, along with the time embedding $s$. However, for the inference step, $\mathbf{w}_S$ and $\mathbf{b}_S$ are both drawn from independent Gaussian distributions, and hence, the learned free energy and dual dissipation potential densities do not depend on either the boundary conditions or process variables.

To ensure that the model outputs strictly follow the governing equations based on the input data, we need to condition the DDPM on the inputs. There are two primary approaches for conditioning a diffusion model, direct conditioning and guidance \cite{croitoru2023diffusion}. Direct conditioning integrates the conditioning information directly, either explicitly or through positional embedding \cite{nichol2021improved, vaswani2017attention, rombach2022high}, into the model during training. This method allows the model to learn the explicit mapping between the input conditions and the target, which is used for this study. 

\subsection{Uncertainty quantification in EVODMs}
\label{sec: epinet}
As mentioned in Section~\ref{sec: intro}, the computational cost of the inference for diffusion models is notably high. This limitation significantly reduces the efficiency of using conditional diffusion models for UQ as one needs to generate thousands of samples to obtain well-calibrated uncertainty estimates. Despite extensive efforts to accelerate the inference process \cite{nichol2021improved, song2020denoising, liu2022pseudo, lu2022dpm, salimans2022progressive}, the generation capabilities of conditional diffusion models have been constrained by the quality-diversity tradeoff, a critical topic in the field of generative modeling \cite{dhariwal2021diffusion, ho2022classifier, rombach2022high, saharia2022photorealistic}. When guidance techniques are employed for conditioning, balancing quality and diversity typically involves tuning the guidance strength \cite{ho2022classifier}. 
However, even with such tuning, the reliability of quantified uncertainty cannot be guaranteed and may compromise the accuracy of the trained diffusion models.

To address these limitations, we adopt an explicit conditioning approach over guidance techniques, ensuring the integrity of the trained models. Furthermore, to maintain model accuracy while reducing computational costs for UQ, an Epinet \cite{osband2023epistemic}, introduced in Section~\ref{sec: enn}, is integrated into our framework and is trained with a physics-informed loss function.

We define the reference distribution of the epistemic index as a standard Gaussian with dimension $d_\Gamma$. This dimension is user-defined and the specific choices used in this work for the numerical examples can be found in Appendix~\ref{appendix: train details}. The learnable network $\sigma_{\eta}^{L}$ takes $h_\zeta(x)$ and index $\gamma$ as inputs, where $h_{\zeta}(x)$ is the concatenation of input $x$ and the features extracted from the last hidden layer of the base network. The learnable network is expressed as 
\begin{equation}
\sigma_\eta^L\left(\mathrm{sg}\left[h_\zeta(x)\right], \gamma\right)=\textit{NN}_\eta\left(\mathrm{sg}\left[h_\zeta(x)\right], \gamma\right)^T \gamma,    
\end{equation}
where $\textit{NN}_\eta$ denotes a neural network parameterized by $\eta$. Its output lies in $\mathbb{R}^{d_\Gamma \times d}$, with $d$ is the dimensionality of the target output. In our framework, the target output corresponds to the free energy or dual dissipation potential densities, implying that $d=1$.
For the fixed prior network $\sigma^P$, we consider an ensemble of $d_\Gamma$ networks, where each network is a small multi-layer perceptron (MLP). Each MLP takes $x$ as input. Let $p^i(x) \in \mathbb{R}^d$ represent the output of the $i^{\text {th}}$ MLP in the ensemble. The prior network can be formulated as 
\begin{equation}
\sigma^P\left(\mathrm{sg}\left[h_\zeta(x)\right], \gamma\right)=\kappa \sum_{i=1}^{d_\Gamma} p^i(x) \gamma_i,
\end{equation}
where $\kappa$ is a scaling factor and $\gamma_i$ is the $i^{\text {th }}$ component of the epistemic index $\gamma$. For the prior network, we do not consider any explicit dependency on $\mathrm{sg}\left[h_\zeta(x)\right]$, consistently with \cite{osband2023epistemic}  Appendix G.3. With that, we simplify the notation to 
\begin{equation}
\sigma^P\left(x, \gamma\right)=\kappa \sum_{i=1}^{d_\Gamma} p^i(x) \gamma_i.
\end{equation}
By combining the base network and the Epinet, Equation~\eqref{eq: enn equation} becomes
\begin{equation}
\label{eq: full enn equation}
y_\theta(x, \gamma)=\mu_\zeta(x)+\sigma_\eta^L\left(\mathrm{sg}\left[h_\zeta(x)\right], \gamma\right)+\sigma^P\left(x, \gamma\right).
\end{equation}

In addition to the low computational cost of ENNs, another key advantage of using ENNs for UQ is their flexibility in training: the Epinet can be trained either concurrently with the base network or after the base network has been fully trained \cite{osband2023epistemic, guilhoto2024composite}. In this study, to optimize computational efficiency, we choose the latter approach, training the Epinets after completing the training of the base network (DDPM). During this phase, the predictions made by the base network are used as the ground truth for the training of Epinets. These predictions provide the denoised target data which are assumed to eliminate the aleatoric uncertainty in the predictive uncertainty and allow us to approximate the epistemic uncertainty via Epinets. This training strategy aligns with the concept of knowledge distillation \cite{hinton2015distilling, gou2021knowledge}, where the base network (the "teacher" model) generates predictions that are used as target data for the Epinets (the "student" model). Rather than learning directly from the original noisy data, the student model learns from the teacher's refined predictions. Our results show that this approach significantly accelerates the convergence of Epinets while preserving prediction accuracy. 
By following this approach, the learned $f$ and $\phi$ (with $\psi$ recovered from $\phi$) become available and could, in principle, serve directly as targets for training the Epinets. However, to guarantee physical consistency, we adopt the $l^2$ norm of the residual of the governing equations derived from Onsager's variational principle as the loss functions to train the Epinets. In particular, the PDE loss term is defined as 
\begin{equation}
\label{eq: epinet loss fn}
\mathcal{L}^{\text{PDE}}_{\text{Epinet}} = \frac{1}{\left|\Gamma_f\right|\left|\Gamma_\phi\right|} \sum_{\boldsymbol{\gamma}_f \in \Gamma_f} \sum_{\boldsymbol{\gamma}_\phi \in \Gamma_\phi} \frac{1}{\left|\mathcal{D}_{\mathrm{PDE}}^{\text{Epinet}}\right|} \sum_{\left(\mathbf{z}, \mathbf{w}_{\boldsymbol{\zeta}_\phi}\right) \in \mathcal{D}_{\mathrm{PDE}}^{\text{Epinet}}}\left(\frac{\partial \phi\left(\mathbf{g}\left(\mathbf{z}, \boldsymbol{\gamma}_f ; \boldsymbol{\theta}_f\right), \boldsymbol{\gamma}_\phi ; \boldsymbol{\theta}_\phi\right)}{\partial \mathbf{g}}-\mathbf{w}_{\boldsymbol{\zeta}_\phi}\right)^2, 
\end{equation}
and, when traction boundary conditions are available, then the boundary loss term is given by
\begin{equation}
\nonumber\\
\mathcal{L}^{\text{BCs}}_{\text{Epinet}} = \frac{1}{\left|\Gamma_f\right|} \sum_{\boldsymbol{\gamma}_f \in \Gamma_f} \frac{1}{\left|\mathcal{D}_{\mathrm{BCs}}^{\mathrm{Epinet}}\right|} \sum_{\left(\mathbf{z}, \mathbf{b}_{\boldsymbol{\zeta}_f}\right) \in \mathcal{D}_{\mathrm{BCs}}^{\text{Epinet}}}\left(\frac{\partial f\left(\mathbf{z}, \boldsymbol{\gamma}_f; \boldsymbol{\theta}_f\right)}{\partial \mathbf{z}}-\mathbf{b}_{\boldsymbol{\zeta}_f}\right)^2.
\end{equation}
The overall loss function for training the Epinets is then simply the sum $\mathcal{L}_{\text{Epinet}} = \mathcal{L}^{\text{PDE}}_{\text{Epinet}} + \mathcal{L}^{\text{BCs}}_{\text{Epinet}}$, with no additional loss weights applied (i.e., they can be seen as set to $1$ in this work). The term $\mathcal{L}^{\text{BCs}}_{\text{Epinet}}$ is dropped in the absence of boundary data. This physics-informed loss function for the Epinet requires the evaluation of derivatives of the ENN outputs $y$ (which represents $f$ or $\phi$) with respect to input $x$ (such as $\mathbf{z}$ or $\mathbf{g}$). These derivatives can be obtained using the chain rule as follows
\begin{equation}
\begin{aligned}
\label{eq: full enn der equation}
\frac{\partial y_\theta(x, \gamma)}{\partial x} & =\frac{\partial \mu_\zeta(x)}{\partial x}+\frac{\partial}{\partial x}\left(\sigma_\eta^L\left(h_\zeta(x), \gamma\right)+\sigma^P(x, \gamma)\right) \\
& =\frac{\partial \mu_\zeta(x)}{\partial x}+\frac{\partial N N_\eta\left(h_\zeta(x), \gamma\right)^{\top}}{\partial h_\zeta(x)} \frac{\partial h_\zeta(x)}{\partial x} \gamma+\kappa \sum_{i=1}^{D_{\Gamma}} \frac{\partial p^i(x)}{\partial x} \gamma_i,
\end{aligned}
\end{equation}
where $\kappa \sum_{i=1}^{D_{\Gamma}} \frac{\partial p^i(x)}{\partial x} \gamma_i$ can be computed prior to the training of the Epinet as the prior network is fixed after initialization. Since we train the base network and the Epinet sequentially, $\frac{\partial \mu_\zeta(x)}{\partial x}$ and $\frac{\partial h_\zeta(x)}{\partial x}$ can be directly obtained from the inference step of the base network. Namely, instead of relying on automatic differentiation to compute these derivatives implicitly and iteratively during training, their explicit form can be derived and computed only once during the inference stage of the base network which can significantly reduce the computational cost. Eventually, only the term $\frac{\partial N N_\eta\left(h_\zeta(x), \gamma\right)^{\top}}{\partial h_\zeta(x)}$ is learned during the training of the Epinet. Note that we omit the $\mathrm{sg}[\cdot]$ as the Epinet is not trained jointly with the base network, meaning that the trained parameters in the base network remain unaffected by the training of Epinet. If the second derivative of $y_\theta$ with respect to $x$ is needed, it can be obtained in a similar fashion. For the simple case of $x\in\mathbb{R}^1$, it reads
\begin{equation}
\begin{aligned}
\frac{\partial^2 y_\theta(x, \gamma)}{\partial x^2}= & \frac{\partial^2 \mu_\zeta(x)}{\partial x^2}+\frac{d}{dx}\left(\frac{\partial \sigma_\eta^L\left(h_\zeta(x), \gamma\right)}{\partial h_\zeta(x)} \frac{\partial h_\zeta(x)}{\partial x}\gamma\right)+\frac{d}{d x}\left(\kappa \sum_{i=1}^{D_{\Gamma}} \frac{\partial p^i(x)}{\partial x} \gamma_i\right) \\
= & \frac{\partial^2 \mu_\zeta(x)}{\partial x^2}+\left[\frac{\partial^2 \sigma_\eta^L\left(h_\zeta(x), \gamma\right)}{\partial h_\zeta(x)^2}\left(\frac{\partial h_\zeta(x)}{\partial x}\right)^2+\frac{\partial \sigma_\eta^L\left(h_\zeta(x), \gamma\right)}{\partial h_\zeta(x)} \frac{\partial^2 h_\zeta(x)}{\partial x^2}\right]\gamma+\kappa \sum_{i=1}^{D_{\Gamma}} \frac{\partial^2 p^i(x)}{\partial x^2} \gamma_i.
\end{aligned}
\end{equation}
In the SSEP example in Section~\ref{sec: sep}, this equation is used to compute the second derivative of the free energy, needed for validation.

Overall, the physics is encoded within the loss functions of the Epinets. By leveraging sequential training and knowledge distillation during the training of Epinets, the model achieves faster convergence and improved stability. These design features allow the Epinets to serve as a robust and efficient component within the EVODMs framework for quantifying epistemic uncertainty.

\subsection{Evaluation metric}
\label{sec: error and evaluation}
To evaluate the error in the mean predictions, we use the relative $l_2$ error (RL2E) averaged over the testing dataset $\mathcal{D}_{\text {Test}}$, defined as
\begin{equation}
\mathrm{RL} 2 \mathrm{E} = \sqrt{\frac{\sum_{i=1}^{N_{\text {Test}}}\left(\bar{\mu}\left(x_i\right)-y_i\right)^2}{\sum_{i=1}^{N_{\text {Test}}} y_i^2}},
\end{equation}
where $\bar{\mu}\left(x_i\right)$ is the mean prediction for the testing data $x_i$ over the samples generated by the trained EVODM. $y_i$ is the corresponding reference value for $x_i$. $N_{\text {Test}}$ is the total number of data points in the testing dataset $\mathcal{D}_{\text {Test}}$.

\subsection{Hardware and implementation}
All the experiments discussed below are run on a single NVIDIA RTX A6000 GPU. Pytorch is the main Python library for implementing  EVODMs proposed in this paper. Standard libraries such as Numpy, Scipy, Pandas, JAX, and Matplotlib are used for data generation, data preprocessing, and plotting results.

\section{Numerical experiments}
\label{sec: experiments}
Here, we demonstrate the capability of EVODMs to learn the free energy and dissipation potential densities with quantified epistemic uncertainties for systems with physical noise. Toward this goal, we will consider the stochastic version of two of the physical examples used in \cite{huang2022variational} and compare the results of EVODMs with VONNs.

\subsection{Phase transformation of coiled-coil proteins}
\label{sec: phase transformation}
The coiled-coil motif, found in approximately $10 \%$ of proteins, consists of $\alpha$-helices intertwined with one another \cite{rose2004scaffolds,torres2019combined}. An interesting feature of this structure is its capability of undergoing a phase transition from the coiled state to an unfolded state in order to sustain large deformations without breaking, of importance for their biological function. This mechanical behavior can be modeled, in a coarse-grained fashion, as a phase transforming rod characterized by a double-well free energy and a dissipation potential to describe the hydrodynamic interaction with the solvent. 

In this example, we will generate noisy data by performing overdamped Langevin dynamic simulations of a one-dimensional rod, based on the non-convex free energy and quadratic dissipation potential estimated by \cite{torres2019combined} via molecular dynamics simulations, and aim at learning such potentials with quantified uncertainty. 

\subsubsection{Model description}
\label{sec: phase trans model description}
Following \cite{torres2019combined} and \cite{huang2022variational}, the protein is modeled as a one-dimensional rod of length $L$, with free energy density $f(\varepsilon)$ and dissipation potential density $\psi(v)=\frac{1}{2}\eta v^2$, where $\varepsilon$ and $v$ are the local strain and velocity, respectively, and $\eta$ is the viscosity. Considering that inertial and body forces are negligible, the equilibrium within the domain and with external tractions $\bar{t}$ can be written, in the deterministic limit, as
\begin{align}
\label{eq: phase_trans}
    \frac{\partial f'(\varepsilon(X,t))}{\partial X} &= \psi'(v(X,t)), \\ 
\label{eq: phase_trans_traction}
    \bar{t} &= f'(\varepsilon(L,t)).
\end{align}

This problem is characterized by a single state variable $z=u$, the displacement field, and a single process variable $w=v$, the velocity field. Using the dual dissipation potential density $\phi(g)=\frac{1}{2}\frac{g^2}{\eta}$, these equations can be rewritten as (see Section \ref{sec: vonns})
\begin{align}
\label{eq: lt_phase_trans}
    v(X,t) &= \phi'(g(\varepsilon(X,t))), \\ 
    \bar{t} &= f'(\varepsilon(L,t)),
\end{align}
which are the physical constraints imposed in the physics-informed loss functions shown in Equation~\eqref{eq: phase trans base loss function}. The same physical constraints are enforced in the training of the Epinets as described in Equation~\eqref{eq: phase trans epinet loss fn}. Both loss functions can be found in Section~\ref{sec: phase trans nn training}.

\subsubsection{Data generation}
\label{sec: phase trans data generation}
\begin{figure}[!htbp]
    \centering
    \includegraphics[width=1.0\textwidth]{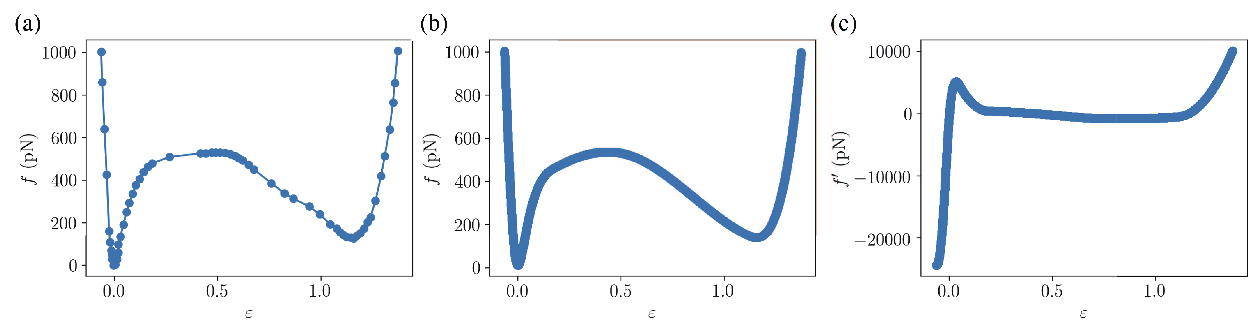}
    \caption{Free energy density of \cite{torres2019combined}, generated from molecular dynamics simulations. (a) Raw data sampled from the reference. (b) Free energy and (c) its derivative, interpolated from (a) at 50,000 equispaced strain values using B-splines. Copyright \copyright Elsevier 2022. Reprinted with permission.}
    \label{fig: protein_ref_data}
\end{figure}

In order to generate noisy data, we consider the stochastic version of Equation~\eqref{eq: phase_trans}, namely,
\begin{equation}\label{Eq:protein_eq}
v(X, t)=\frac{1}{\eta} \frac{\partial f^{\prime}(\varepsilon(X, t))}{\partial X}+\sqrt{\frac{2 k_B T}{\eta}} \dot{\omega}_{x,t},
\end{equation}
where we have included the Brownian noise induced by the solvent. Here, $k_B$ is the Boltzmann constant, $T$ is the temperature, and $\dot{\omega}_{x,t}$ is a space-time white noise satisfying $\langle \dot{\omega}_{x,t}\rangle =0$, and $\langle \dot{\omega}_{x,t}\dot{\omega}_{x',t'}\rangle=\delta(t-t')\delta(x-x')$, where $\langle \rangle$ denotes the ensemble average. We use the free energy data obtained by \cite{torres2019combined} via molecular dynamics simulations, as well as their estimated value for the viscosity, $\eta=8$ pN ns nm$^{-2}$. The free energy is represented in Figure~\ref{fig: protein_ref_data}, where the raw data is shown in panel (a) and the interpolated values, following \cite{huang2022variational}, and its derivative are shown in panels (b) and (c), respectively.

\begin{figure}[!htbp]
    \centering
    \includegraphics[width=0.7\textwidth]{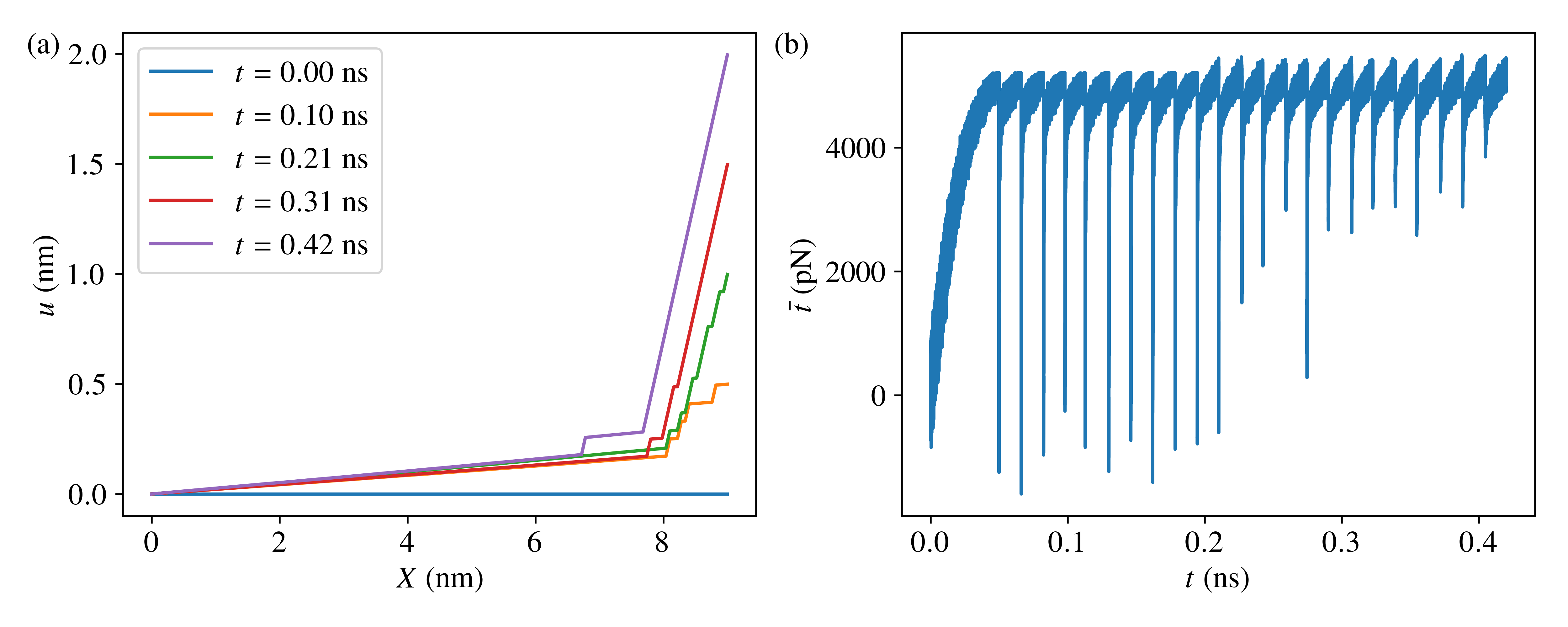}
    \caption{Data generated for the coil-coiled protein model exhibiting a phase transformation. (a) Displacement field at various instants of time. (b) Temporal evolution of the external traction.}
    \label{fig: u_traction_phasetrans}
\end{figure}

The simulated rod is of length $L=9$ nm, fixed at $X=0$, and pulled at $X=L$ with a constant velocity $v_p/2=4.75$ m/s. 
The SDE is discretized using a finite difference scheme to approximate the spatial derivatives and the Euler-Maruyama method for temporal integration. Both space and time are uniformly discretized, with $N_X+1$ nodes spaced by $\Delta X=L / N_X=9/150$ ns, and $N_T$ time steps of size $\Delta t=3 \times 10^{-8}$. The thermal energy is set to $k_B T=10^{14} $ pN nm, and the total simulation time is $0.42$ ns.
The displacement field $u$ and the external traction $\bar{t}$ at $X=L$ are recorded, and these are shown in Figure~\ref{fig: u_traction_phasetrans}. As expected, the results are noisy, and characterized by a phase transforming front that starts at the pulling end and propagates through the system. 

\subsubsection{Data preprocessing}
\label{sec: phase trans data preprocess}
Due to the small timestep, the raw dataset is extremely large, with the total number of data points approaching 2.1 billion. Additionally, this raw dataset is also highly imbalanced due to the double-well structure of the free energy. To improve the training efficiency, we adopt the data pre-processing strategy of \cite{huang2022variational} to sample a certain number of data points from the raw dataset. More specifically, for the free energy density $f(\varepsilon)$, which is primarily determined from the boundary conditions, we sample the boundary strain $\varepsilon^k_{N_X}$ as uniformly as possible and then collect the corresponding traction data $\bar{t}^k$. Since the dual dissipation potential density $\phi(g)$ is learned from the PDE, we sample the strain data $\varepsilon_i^k$, with $i=1, \ldots, N_X-1$ and $k=1, \dots, N_T-1$, to be as uniformly distributed as possible and collect the corresponding velocity data $v_i^k$ and the neighboring strain data $\varepsilon_{i+1}^k$ which is later used in the finite difference method. 

This preprocessing step helps us to construct a relatively balanced dataset across the input space. In total, we sample $1000$ data points for the BC and another $1000$ for the PDE. We randomly select $80 \%$ of the data for training and reserve the remaining $20 \%$ for validation.

\subsubsection{Training and loss functions}
\label{sec: phase trans nn training}
In this example, we use an INN to learn the free energy density $f(\varepsilon,\bar{t}_s,s)$ and a PICINN to learn the dual dissipation potential density $\phi(g,v_s,s)$, which is later used to recover $\psi(v)$ via Legendre transform of the inferred $\phi$. The input to the PICINN, $g$, is computed from 
\begin{equation}
g=\frac{\partial \psi}{\partial v}=\frac{\partial f^{\prime}(\varepsilon(X, t), \bar{t}_S,s)}{\partial X},
\end{equation}
using the finite difference method, i.e.,
\begin{equation} \label{Eq:g_calculation}
g_i^k = \frac{f^{\prime}\left(\varepsilon^{k}_{i+1} ,\bar{t}_S,s; \boldsymbol{\zeta}_f\right)-f^{\prime}\left(\varepsilon^{k}_i,\bar{t}_S,s ; \boldsymbol{\zeta}_f\right)}{\Delta X}, \quad i=1, \dots N_{X}-1 \text{ and } k=1, \dots, N_{T}-1.     
\end{equation}
Here, $\bar{t}_S$ is drawn from a unit Gaussian distribution. The loss function for training the base network is defined as
\begin{equation}
\mathcal{L}_{\text{Base}}=\lambda_{\text{PDE}} \cdot \mathcal{L}_{\text{PDE}}+\lambda_{\text{BC}} \cdot \mathcal{L}_{\text{BC}},
\end{equation}
with
\begin{align}
\label{eq: phase trans base loss function}
\mathcal{L}_{\text{PDE}} &= \mathbb{E}_{\substack{\left(\varepsilon_i^k, v_i^k\right) \sim \mathcal{D}_{\text {PDE}}, \\
s \sim \text{Uniform}[1,S]}}\left[\left|v_i^k-\frac{\partial \phi(v_{i,s}^k, g_i^k, s; \boldsymbol{\zeta}_{\phi})}{\partial g_i^k}\right|^2\right], \\ \nonumber
\quad \mathcal{L}_{\text{BC}}&=\mathbb{E}_{\substack{\left(\varepsilon_{N_X}^k, \bar{t}^k\right) \sim \mathcal{D}_{\text{BC}}, \\
s \sim \text{Uniform}[1,S]}}\left[\left|\bar{t}^k-\frac{\partial f\left(\bar{t}^k_s, \varepsilon_{N_X}^k, s; \boldsymbol{\zeta}_f\right)}{\partial \varepsilon_{N_X}^k}\right|^2\right].
\end{align}
The input dataset for the PDE, $\mathcal{D}_{\text{PDE}}$, is $\left(\varepsilon_i^k, \varepsilon_{i+1}^k, v_i^k\right)$ with $i=1, \ldots, N_X-1$ and $k=0, \ldots, N_T-1$. Similarly, the input dataset for the BC, $\mathcal{D}_{\text{BC}}$, is $\left(\varepsilon_{N_X}^k, \bar{t}^k\right)$. Meanwhile, $v_{i,s}^k$ and $\bar{t}^k_s$ represents the corrupted version of $v_i^k$ and $\bar{t}^k$, respectively, at diffusion time step $s$. In the training, we set both of the loss weights to be $1$.

We remark that $g$, since it is computed via Eq.~\eqref{Eq:g_calculation}, is not directly accessible prior to training, making its normalization infeasible, and thus affecting the training stability in this example. To solve this issue, we implement the following two-phase training protocol for this example. In the first phase, the PDE component of the loss function is frozen, and only $\zeta_f$ is updated using the residual of the BC equation. At the end of this phase, the mean and standard deviation of $g$ are estimated and stored for normalizing $g$ before entering the second phase. In the second phase, the PDE component is unfrozen, and the full loss function is trained, updating both $\zeta_f$ and $\zeta_\phi$. Despite its simplicity, this protocol significantly improves the numerical stability, training efficiency, and robustness of the model. 

After training the base network, we train small Epinets with physics-informed loss functions defined as
\begin{align}
\label{eq: phase trans epinet loss fn}
\mathcal{L}_{\text{Epinet}} = & \frac{1}{\left|\Gamma_f\right|\left|\Gamma_\phi\right|} \sum_{\boldsymbol{\gamma}_f \in \Gamma_f} \sum_{\boldsymbol{\gamma}_\phi \in \Gamma_\phi} \frac{1}{\left|\mathcal{D}_{\mathrm{PDE}}^{\text{Epinet}}\right|} \sum_{\left(\varepsilon_i^k, v_{\boldsymbol{\zeta}_\phi}\right) \in \mathcal{D}_{\mathrm{PDE}}^{\text{Epinet}}}\left(\frac{\partial \phi\left(g_i^k\left(\varepsilon_i^k, \boldsymbol{\gamma}_f ; \boldsymbol{\theta}_f\right), \boldsymbol{\gamma}_\phi ; \boldsymbol{\theta}_\phi\right)}{\partial g_i^k}-v_{\boldsymbol{\zeta}_\phi}\right)^2 \nonumber\\
&\quad + \frac{1}{\left|\Gamma_f\right|} \sum_{\boldsymbol{\gamma}_f \in \Gamma_f} \frac{1}{\left|\mathcal{D}_{\mathrm{BC}}^{\mathrm{Epinet}}\right|} \sum_{\left(\varepsilon_{N_X}^k, \bar{t}_{\boldsymbol{\zeta}_f}\right) \in \mathcal{D}_{\mathrm{BC}}^{\text{Epinet}}}\left(\frac{\partial f\left(\varepsilon_{N_X}^k, \boldsymbol{\gamma}_f; \boldsymbol{\theta}_f\right)}{\partial \varepsilon_{N_X}^k}-\bar{t}_{\boldsymbol{\zeta}_f}\right)^2,
\end{align}
where $\mathcal{D}^{\text{Epinet}}_{\text{PDE}}$ and $\mathcal{D}^{\text{Epinet}}_{\text{BC}}$ are the training datasets used to train the Epinets, respectively. The predictions from the base networks $v_{\boldsymbol{\zeta}_\phi}$ and $\bar{t}_{\boldsymbol{\zeta}_f}$ are used as the target data in the loss function. Although the complete trainable parameters $\boldsymbol{\theta} = (\boldsymbol{\zeta}, \boldsymbol{\eta})$ are used here, the training of the Epinets updates only $\boldsymbol{\eta}$, leaving $\boldsymbol{\zeta}$ from the base network fixed. Full details of the neural network architectures and the hyperparameters used for this example can be found in Appendix~\ref{appendix: train details}.

\subsubsection{Results}
\label{sec: phase trans results}
\begin{figure}[!htbp]
    \centering  \includegraphics[width=0.9\textwidth]{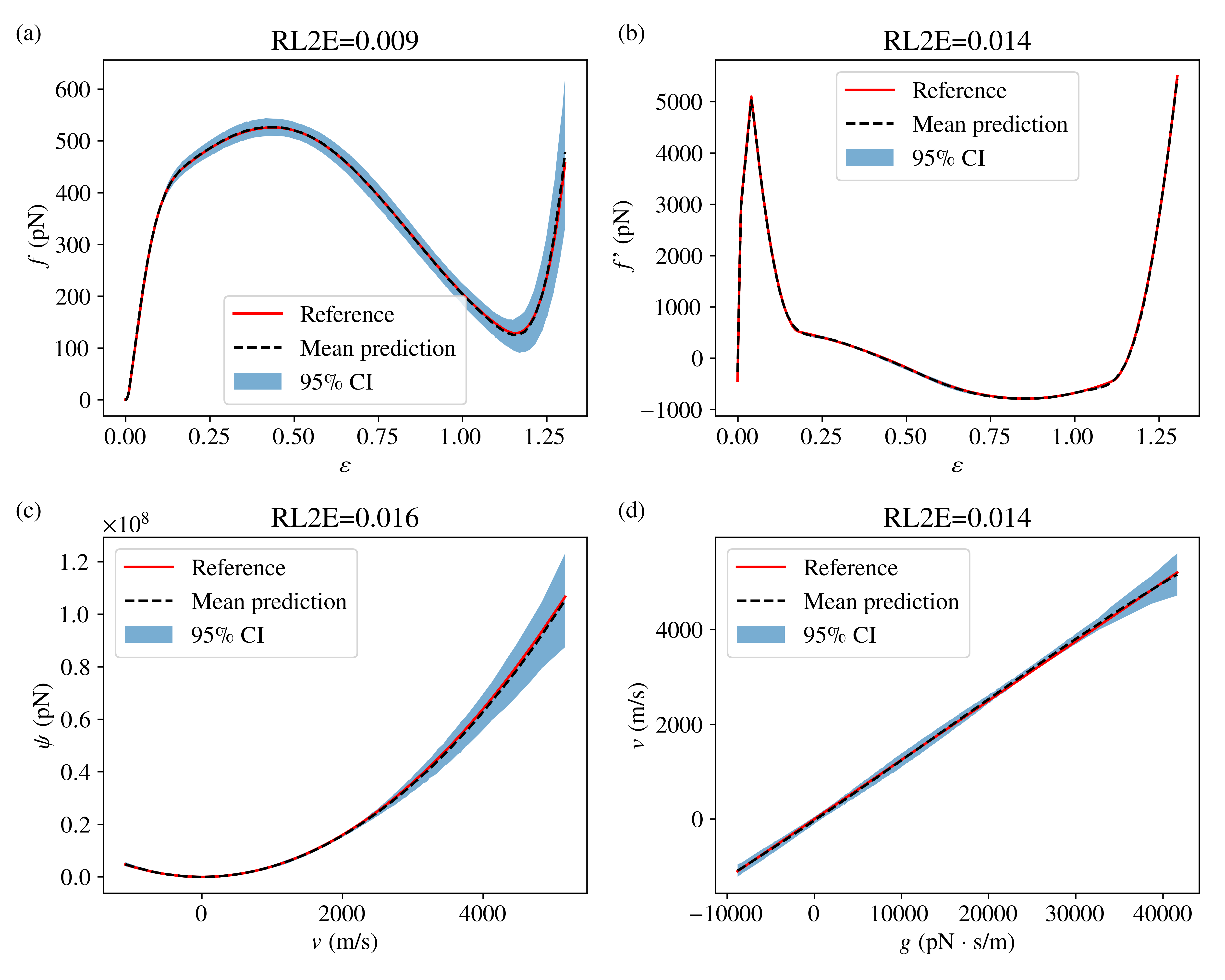}
    \caption{Results for (a) the free energy density $f(\varepsilon)$, (b) its derivative $f'(\varepsilon)$, (c) the dissipation potential density $\psi(v)$, and (d) its derivative $\psi'(v)$ for a one-dimensional phase transforming rod. The solid red lines represent the reference values, the dashed black lines are the mean predictions given by EVODMs, and the shaded blue regions denote the epistemic uncertainties quantified by EVODMs.}
    \label{fig: phasetrans results}
\end{figure}

To evaluate the mean predictions and quantify epistemic uncertainties, 2000 samples are generated for each QoI using the trained EVODM. The results, presented in Figure~\ref{fig: phasetrans results}, demonstrate an excellent agreement with the reference values for both potentials, the free energy density $f(\varepsilon)$, and the dissipation potential density $\psi(v)$ (recovered from its learned dual $\phi$), and their respective derivatives. The relative $l_2$ errors for $f(\varepsilon), f^{\prime}(\varepsilon), \psi(v)$, and $\psi^{\prime}(v)$ are $0.9 \%, 1.4 \%, 1.6 \%$, and $1.4 \%$, respectively. Additionally, each figure includes the $95 \%$ confidence interval (CI) to represent the quantified epistemic uncertainty. We can clearly observe how epistemic uncertainties propagate from the derivatives back to the corresponding densities. For comparison, the same dataset is used to train and validate VONNs. The results, shown in Figure~\ref{fig: phase_trans_vonns_results}, yield relative $l_2$ errors for $f(\varepsilon), f^{\prime}(\varepsilon), \psi(v)$, and $\psi^{\prime}(v)$ of $1.0 \%, 2.5 \%, 34.7 \%$, and $34.3 \%$, respectively. While VONNs can learn $f(\varepsilon)$ and $f^{\prime}(\varepsilon)$ with relatively low errors, they completely fail to produce accurate results for $\psi(v)$ and $\psi^{\prime}(v)$ due to the inherent stochasticity in the data. Furthermore, the deterministic nature of the vanilla VONNs used in this comparison prevents them from quantifying the epistemic uncertainty.

\subsection{Symmetric Simple Exclusion Process (SSEP)}
\label{sec: sep}
The simple exclusion process is a lattice process, where particles sit on lattice sites and randomly attempt to move to a neighboring site at a constant rate (here chosen to be 1 without loss of generality). Due to the exclusion rule, if the neighboring site is already occupied, the particle's movement is rejected, and it is set to remain in its current site. This restriction ensures that each site can hold no more than one particle at any time, leading to dynamics in which particle interactions arise solely from the competition for site occupancy. Here, we consider the one-dimensional case, where a particle has an equal probability of moving to any unoccupied neighboring site (i.e., the symmetric case). 
For more details about SSEP, see \cite{kipnis2013scaling, embacher2018computing, huang2020particle}. 

\subsubsection{Model description}
In the limit of infinitely many particles, the density evolves according to the diffusion equation
\begin{equation} 
\label{Eq:diffusioneq}
\frac{\partial \rho}{\partial t}=D \nabla^2 \rho,
\end{equation}
where $D=1/2$ is the diffusion coefficient. In thermodynamic form, this equation can be written as
\begin{equation}
    \frac{\partial \rho}{\partial t} = \nabla \cdot \left( m(\rho) \nabla \frac{\delta \mathcal{F}}{\delta \rho} \right)
\end{equation}
where $\mathcal{F}=\int_{\Omega} \beta^{-1} \left(\rho \log \rho + (1-\rho)\log(1-\rho\right))\, dV$ is the system's free energy with $\beta$ the inverse temperature, and $m(\rho) = \frac{\beta}{2}\rho(1-\rho)$ is the mobility coefficient. We note that while the free energy and mobility are uniquely determined for the given particle process, these cannot be uniquely determined from Eq.~\eqref{Eq:diffusioneq}, i.e., multiple pairs of free energy and mobility coefficient can give rise to the same dynamics.

This equation can be derived from Onsager's variational principle, using the density field $z=\rho$ as the state variable and the flux $\mathbf{w}=\mathbf{j}$ as the process variable. These two variables are related through the mass conservation equation 
\begin{equation}
\label{eq: diffMassConserv}
\frac{\partial \rho}{\partial t}=-\nabla \cdot \mathbf{j} .
\end{equation}
The constitutive relation $j=-m(\rho)\nabla \frac{\delta \mathcal{F}}{\delta \rho}$ then follows from Onsager's variational principle with the free energy given by $\mathcal{F}[\rho]=\int_{\Omega}f(\rho) \, dV$, 
and the dissipation potential density $\psi(\rho, j)=\frac{1}{2}\frac{j^2}{m(\rho)}$, i.e.
\begin{equation}
\label{eq: diffKinetic}
    \nabla f'(\rho) + \frac{\partial \psi(\rho, \mathbf{j})}{\partial \mathbf{j}} = \mathbf{0}.
\end{equation}
While $f(\rho)$ and $\psi(\rho,j)$ are not unique from the dynamics as noted earlier, the auxiliary function
\begin{equation}
\label{eq: psi_hat_sep_ratio}
    \hat{\psi}(\rho, j) = \frac{\psi(\rho, j)}{f^{\prime \prime}(\rho)}
\end{equation}
is however unique.
For this example, it takes the form
\begin{equation}
    \hat{\psi}(\rho, j) = j^2.
\end{equation}
This auxiliary function serves as a reference for validating the model in Section~\ref{sec: sep results}.

\subsubsection{Data generation}
\begin{figure}[!htbp]
    \centering    \includegraphics[width=1.0\textwidth]{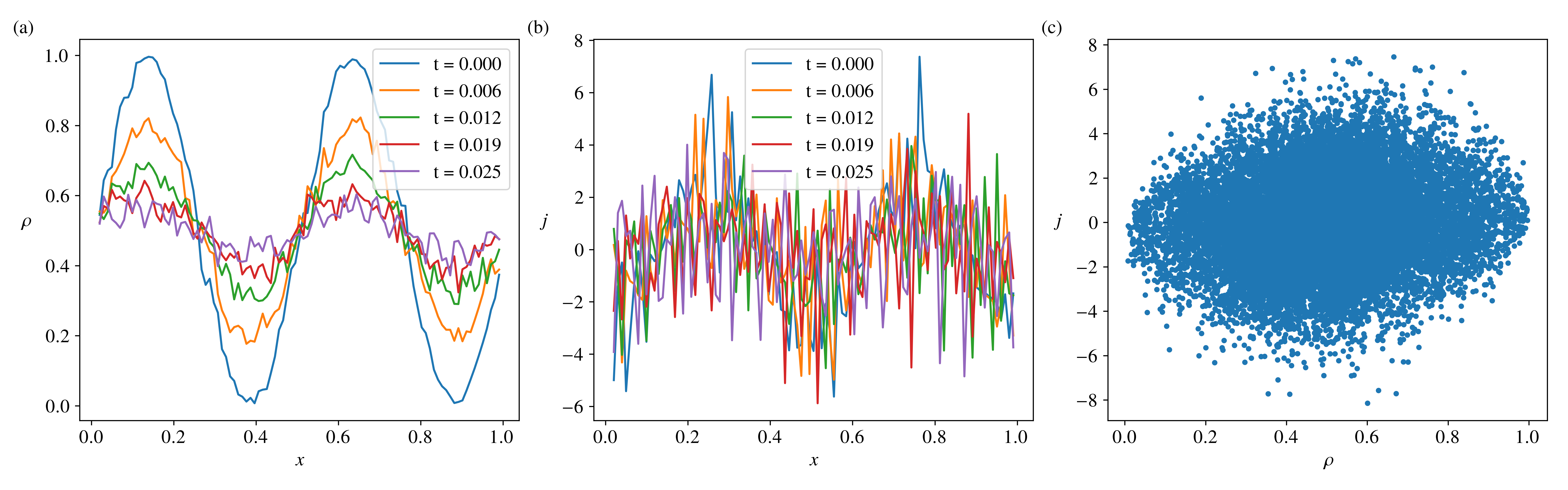}
    \caption{Dataset for the symmetric simple exclusion process. (a) Snapshots of the coarse-grained density field, and (b) the flux field, and (c) input space $\rho-j$ sampled during the simulations.}
    \label{fig: sep_dataset}
\end{figure}

The SSEP considered is one-dimensional with periodic boundary conditions and contains $L=N_{site}=20000$ sites. It is simulated using Kinetic Monte Carlo, in particular, the Bortz-Kalos-Lebowitz (BKL) algorithm \cite{bortz1975new}. Macroscopically, the domain is set to be of length $1$, and it is uniformly discretized into $N_x=100$ elements, each of length $\Delta x = 1/N_x$. The total macroscopic simulation time is $t=0.025$. Macroscopic density information is computed at each node $x_i$ from the particle data at time intervals $\Delta t=1.25 \times 10^{-4}$ as 
\begin{equation}
\rho(x_i, t)=\sum_{X \in\{1, \ldots, L\}} \gamma_i(\frac{X}{L}) \eta\left(t L^2, X\right) \frac{1}{L \Delta x},
\end{equation}
where $\{\gamma_i(x)\}$ are linear finite element shape functions defined as $\gamma_i(x)=\max \left(1-N_x\left|x-x_i\right|, 0\right)$, and $\eta$ represents a lattice configuration ($\eta(T, X)=1$ if there is a particle at time $T$ in site $X$, otherwise, $\eta(T, X)=0$). We note that the microscopic coordinates $(X, T)$ and the macroscopic ones $(x,t)$ are related through the parabolic scaling, $x=\frac{X}{L}$ and $t=\frac{T}{L^2}$ \cite{kipnis2013scaling}. The flux is obtained from the density field at each coarse-grained position over time $\rho_i^k := \rho(x_i, t^k)$ using a finite difference approximation as
\begin{equation}
\label{eq: sep_flux_compute}
    j_{i+\frac{1}{2}}^k = - \frac{1}{2}\frac{\rho_{i+1}^k - \rho_i^k}{\Delta x}.
\end{equation}
Following this setup, the generated dataset of one realization is shown in Figure~\ref{fig: sep_dataset}, where fluctuations in both the density and flux fields are clearly visible.

\subsubsection{Data preprocessing}
\begin{figure}[!htbp]
    \centering
    \includegraphics[width=0.8\textwidth]{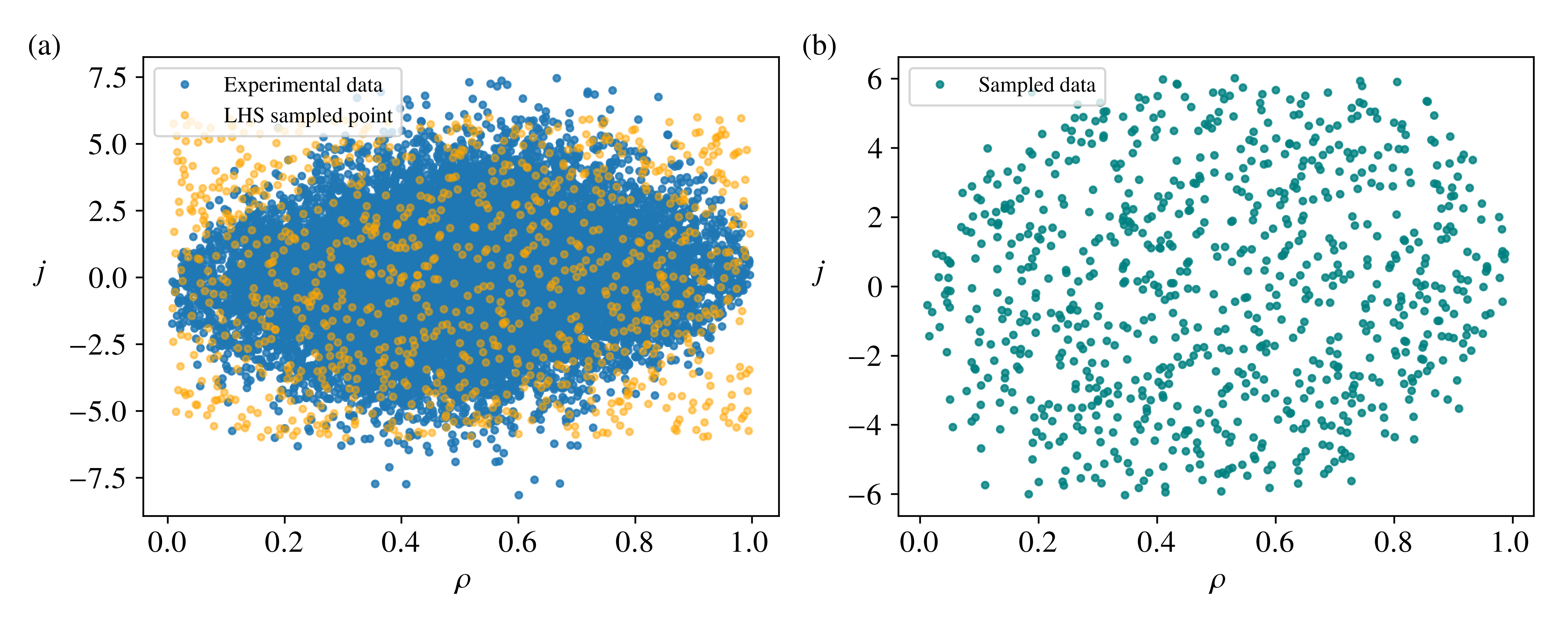}
    \caption{Input data space before and after preprocessing. (a) Numerical experimental data before preprocessing and the query points sampled by LHS that are used for preprocessing. (b) Sampled data after preprocessing.}
    \label{fig: sep_datapreprocess}
\end{figure}
For data preprocessing, we first generate samples in the $\rho-j$ space using Latin hypercube sampling (LHS) \cite{mckay2000comparison} to serve as query points. LHS divides each variable's range into equally probable intervals and ensures one sample per interval. As shown in Figure~\ref{fig: sep_datapreprocess}(a), the minimum and maximum values of $\rho$ are used as the lower and upper bounds for the input variable. Due to the large variation in $j$, we manually set the lower and upper bounds for $j$ to be $-6$ and $6$ with only a small partial of data points disregarded. Using the K-d tree algorithm \cite{bentley1975multidimensional}, we then identify the nearest neighbors from the dataset. To avoid redundancy, we remove any duplicated sampled points. 
The resultant dataset is a more balanced and space-filling one compared to the one generated by uniform grid method used in \cite{huang2022variational}. Overall, we sample $813$ data points shown in Figure~\ref{fig: sep_datapreprocess}(b), and randomly choose $80 \%$ of these sampled points for training, while the remaining data points are used for validation. 

\subsubsection{Training and loss functions}
An important aspect of this example is the nonuniqueness: the free energy and dissipation potential densities cannot be uniquely inferred from the data. To validate the model, we focus on the auxiliary function $\hat{\psi}(\rho, j)$. 
Although a single neural network could, in principle, approximate the auxiliary function, the convexity of $\hat{\psi}$ with respect to $j$ is not guaranteed unless the convexity of $f(\rho)$ is established. Therefore, we use an INN to learn the free energy density $f(\rho,j_s,s)$ and a PICINN to learn the dual dissipation potential density $\phi(g,j_s,s)$, which is used later to recover $\psi$. 

Unlike the first example, the scale of the data in this example is not hindering the training. The two-phase training protocol is thus not employed here, though it could be an option to further enhance the training stability. The loss function used for training the base model is defined as
\begin{equation}
\label{eq: sep base loss function}
\mathcal{L}_{\text{Base}} = \mathcal{L}_{\text{PDE}} = \mathbb{E}_{\substack{\left(\rho_i^k, j_{i+\frac{1}{2}}^k\right) \sim \mathcal{D}_{\text{PDE}}, \\
s \sim \text{Uniform}[1, S]}}\left[\left|j_{i+\frac{1}{2}}^k+\frac{\partial \phi\left(j_{i+\frac{1}{2},s}^k, g_{i+\frac{1}{2}}^k, s; \boldsymbol{\zeta}_{\phi}\right)}{\partial g_{i+\frac{1}{2}}^k}\right|^2\right],
\end{equation}
where $g_{i+\frac{1}{2}}^k$ is obtained via
\begin{equation}
g_{i+\frac{1}{2}}^k = \frac{f^{\prime}\left(\rho^k_{i+1}, j_{S}, s; \boldsymbol{\zeta}_f\right)-f^{\prime}\left(\rho^k_i, j_{S}, s; \boldsymbol{\zeta}_f\right)}{\Delta x}, \quad i=0, \dots N_{x}-1 \text{ and } k=0, \dots, N_{t}.     
\end{equation}
Here, $j_{i+\frac{1}{2},s}^k$ represents the noised version of $j_{i+\frac{1}{2}}^k$ at diffusion time step $s$ obtained from the forward diffusion process and $j_S$ is sampled from a unit Gaussian distribution. Thus, the input dataset for the base network $\mathcal{D}_{\text{PDE}}$ is $\left(\rho_i^k, \rho_{i+1}^k, j_{i+\frac{1}{2}}^k\right)$. The sequential training for the Epinets uses the following loss function 
\begin{equation}
\label{eq: sep_epinet_loss_fn}
\mathcal{L}_{\text{Epinet}} = \frac{1}{\left|\Gamma_f\right|\left|\Gamma_\phi\right|} \sum_{\boldsymbol{\gamma}_f \in \Gamma_f} \sum_{\boldsymbol{\gamma}_\phi \in \Gamma_\phi} \frac{1}{\left|\mathcal{D}_{\mathrm{PDE}}^{\text{Epinet}}\right|} \sum_{\left(\rho_i^k, j_{\boldsymbol{\zeta}_\phi}\right) \in \mathcal{D}_{\mathrm{PDE}}^{\text{Epinet}}}\left(j_{\boldsymbol{\zeta}_\phi} + \frac{\partial \phi\left(\rho_i^k, g_{i+\frac{1}{2}}^k\left(\rho_i^k,\,\boldsymbol{\gamma}_f; \boldsymbol{\theta}_f\right),\,\boldsymbol{\gamma}_\phi; \boldsymbol{\theta}_\phi\right)}{\partial g_{i+\frac{1}{2}}^k}\right)^2,
\end{equation}
where $\mathcal{D}^{\text{Epinet}}_{\text{PDE}}$ is the training dataset for the Epinets, which contains the prediction from the base network $j_{\boldsymbol{\zeta}_\phi}$. For the trainable parameters $\boldsymbol{\theta} = (\boldsymbol{\zeta}, \boldsymbol{\eta})$, the training of the Epinets updates only $\boldsymbol{\eta}$ as discussed before. The architectures of the neural networks and the hyperparameters are provided in detail in Appendix~\ref{appendix: train details}. 

\subsubsection{Results}
\label{sec: sep results}
\begin{figure}[!htbp]
    \centering    \includegraphics[width=1.0\textwidth]{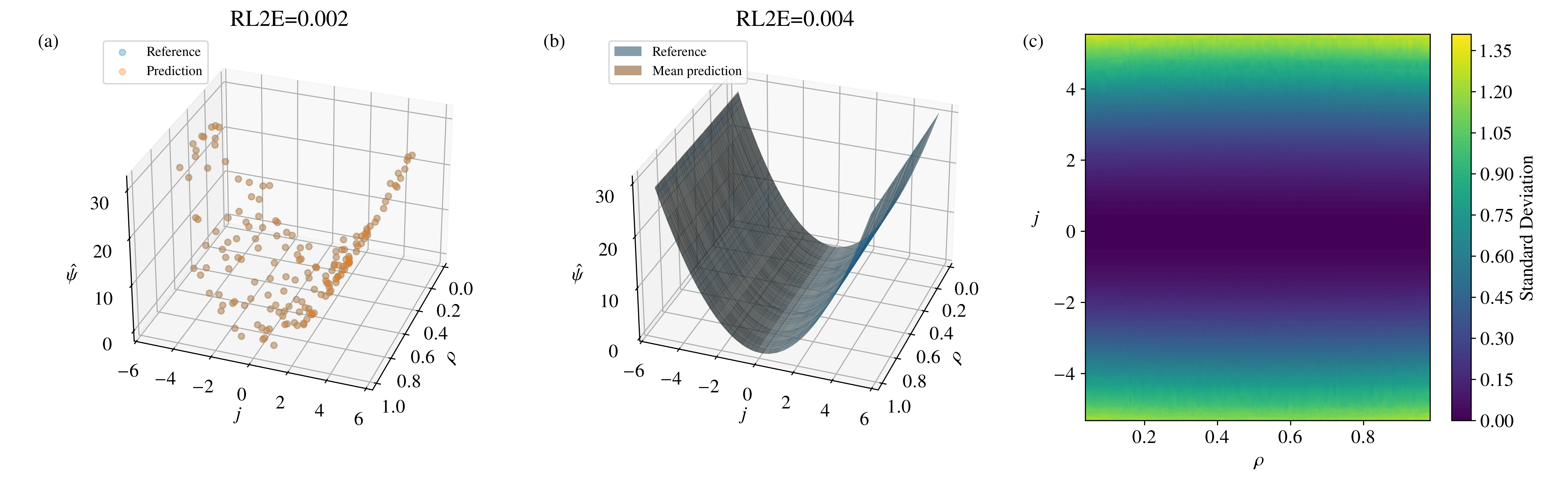}
    \caption{EVODM results for the symmetric simple exclusion process. (a) Predictions for $\hat{\psi}$ generated by the base model of EVODM compared against reference values on the testing dataset; (b) mean predictions for $\hat{\psi}$ obtained by the whole EVODM compared with reference values over the entire surface span by the $\rho-j$ grid; and (c) the standard deviation representing the epistemic uncertainty quantified by the Epinets.}
    \label{fig: sep_evodm_results}
\end{figure}
After the training, we first use the trained base model of the EVODM to make predictions for the testing dataset. The relative $l^2$ error on this testing dataset is low at $0.2 \%$ as shown in Figure~\ref{fig: sep_evodm_results}(a). Later, combining with the trained Epinets, 2000 samples are generated for each point in a $\rho$-$j$ grid as shown in Figure~\ref{fig: sep_evodm_results}(b). These samples are used to compute the mean predictions and quantify the epistemic uncertainties. Although the majority of the surface area is outside the input data space, the relative $l^2$ error of the resulting surface only increases to $0.4 \%$ as presented in Figure~\ref{fig: sep_evodm_results}(b). This indicates that the trained model generalizes well and demonstrates robustness. The quantified epistemic uncertainty (represented by the standard deviation) in Figure~\ref{fig: sep_evodm_results}(c) aligns with the fact that regions with higher uncertainties correspond to areas where fewer or no data points are collected for training. For comparison, the same dataset is used to train and validate VONNs. The results in Figure~\ref{fig: sep_vonns_results}(a) and (b) show that relative $l^2$ errors obtained from VONNs on the testing dataset and the full $\rho$-$j$ grid are $17.3 \%$ and $18.2 \%$, respectively. These results demonstrate the superior ability of EVODMs over VONNs to generate accurate and robust predictions in the presence of noisy data.

\section{Conclusion}
\label{sec: conclusion}
In this paper, we introduce Epistemic Variational Onsager Diffusion Models (EVODMs), a machine learning framework that combines Onsager's variational principle to model non-equilibrium phenomena, with conditional DDPMs from generative modeling and Epinets for uncertainty quantification. 
EVODMs provide an accurate and robust approach for learning free energy and dissipation potential densities from noisy data. By further integrating it with Epinets, EVODMs are also capable of quantifying epistemic uncertainty with minimal computational cost added. Leveraging the variational structure, this framework requires only the state and process variables as inputs and directly learns the unobservable potentials from these observables. Moreover, the governing equations derived from Onsager's variational principle are encoded in the loss functions for physics-informed learning of both the base networks and the Epinets, while the second law of thermodynamics is strongly enforced within the neural network architectures. This ensures the thermodynamic consistency of the learned evolution equations and enhances the framework’s reliability.

Two examples are presented to demonstrate the validity of EVODMs: the phase transformation of a coiled-coil protein, which involves a non-convex free energy density and is modeled via Langevin dynamics, and a particle process (the symmetric simple exclusion process), whose macroscopic dynamics are aimed to be discovered. In both examples, EVODMs exhibit outstanding performance in learning the free energy and dissipation potential densities, demonstrating their robust ability to infer unobservable quantities from noisy observable measurements. Furthermore, EVODMs are able to quantify the associated epistemic uncertainties with minimal additional computational cost, underscoring their reliability and computational efficiency.

Future work includes replacing DDPMs with alternative diffusion models, such as Denoising Diffusion Implicit Models (DDIMs) \cite{song2020denoising}, to further reduce the computational cost and accelerate inference. The proposed framework can also be extended with active learning \cite{cohn1996active,settles2009active}. Finally, we will focus on improving EVODMs to tackle more complex and high-dimensional physics systems.


\appendix
\section{INN for Free Energy Density}
\label{appendix: inn free energy density}
\begin{figure}[!htbp]
    \centering
    \includegraphics[width=0.8\textwidth]{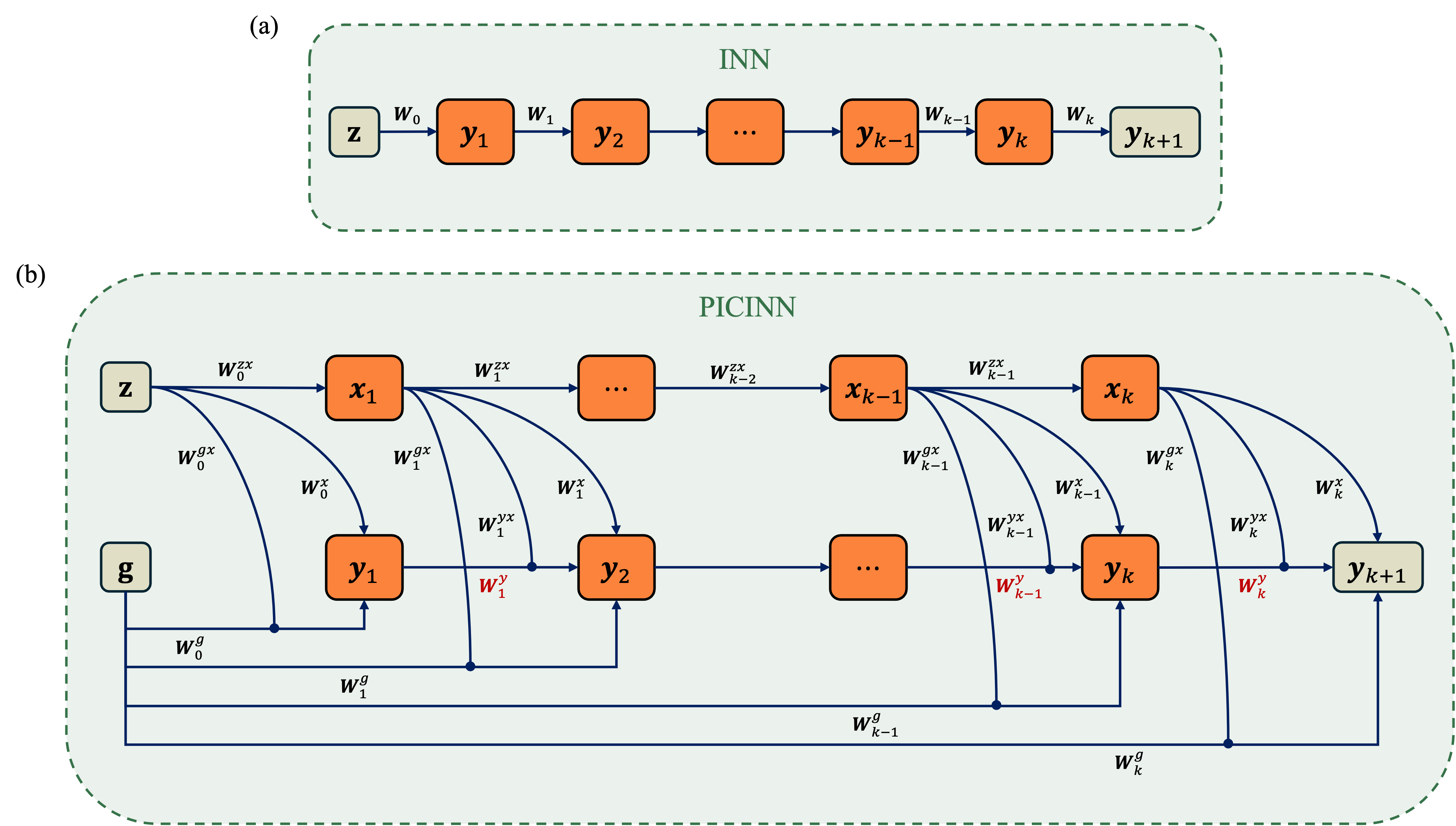}
    \caption{Illustrations of the (a) Integrable Neural Network (INN), and (b) Partially Input Convex Integrable Neural Network (PICINN). For a PICINN, the weights in red $\boldsymbol{W}_i^y$ with $i=1, \cdots, k$ are constrained to be non-negative, while the weights shown in black are standard weights without constraints.}
    \label{fig: inn_picinn scheme}
\end{figure}
In EVODMs, the free energy density $f$ is learned using an INN as illustrated in Figure~\ref{fig: inn_picinn scheme}(a). For each layer in the INN, the outputs $\boldsymbol{y}_{i+1}$ can be expressed as
\begin{equation}
\boldsymbol{y}_{i+1}=a_i\left(\boldsymbol{W}_i \boldsymbol{y}_i+\boldsymbol{b}_i\right), \quad i=0, \ldots k,
\end{equation}
where $\boldsymbol{W}_i,$ and $\boldsymbol{b}_i$ represent the weights and biases of the layer, respectively. $a_i(\cdot)$ denotes the activation function applied to the transformed inputs. The input of the INN is $\boldsymbol{y}_0=\mathbf{z}$.

INNs require that the activation functions $a_i(\cdot)$ used in the hidden layers must be chosen such that their derivatives $a_i^{\prime}(\cdot)$ are themselves common activation functions \cite{teichert2019machine}. In this work, we choose the SoftPlus function as our only activation function for all models and examples. The SoftPlus function can be written as
\begin{equation}
a(x)=\log \left(1+e^x\right),
\end{equation}
and its derivative is given by $a^{\prime}(x)=1 /\left(1+e^{-x}\right)$. This derivative is the logistic function, which is another common activation function.

\section{PICINN for Dissipation Potential Density}
\label{appendix: picinn dissipation potential density}
The dual dissipation potential density $\phi$ is learned by a PICINN. The architecture of a PICINN is shown in Figure~\ref{fig: inn_picinn scheme}(b). The outputs of this can be written mathematically as
\begin{equation}
\label{eq: picinn}
\begin{split}
    & \boldsymbol{x}_{i+1} = a_i^x \left( \boldsymbol{W}_i^{zx} \boldsymbol{x}_i + \boldsymbol{b}_i^{zx} \right),
    \quad \quad 
    i = 0, 1, \dots, k-1 \\
    & \boldsymbol{y}_{i+1} = a_i \left( 
    \widetilde{\boldsymbol{W}}_i^y \left[ \boldsymbol{y}_i \circ 
    a_i^{yx} \left( \boldsymbol{W}_i^{yx} \boldsymbol{x}_i + \boldsymbol{b}_i^{yx} \right) \right]
    + \boldsymbol{W}_i^g \left[ \mathbf{g} \circ 
    \left( \boldsymbol{W}_i^{gx} \boldsymbol{x}_i + \boldsymbol{b}_i^{gx} \right) \right]
    + \boldsymbol{W}_i^x \boldsymbol{x}_i + \boldsymbol{b}_i \right),
    \quad \quad 
    i = 0, 1, \dots, k,
\end{split}
\end{equation}
where $\boldsymbol{x}_{i}$ and $\boldsymbol{y}_{i}$ are the outputs of layer $i$ in the non-convex part and convex part, respectively; $\boldsymbol{W}_i^{zx}$, $\widetilde{\boldsymbol{W}}_i^y$, $\boldsymbol{W}_i^{yx}$, $\boldsymbol{W}_i^g$, $\boldsymbol{W}_i^{gx}$ and $\boldsymbol{W}_i^x$ are the weights matrices associated with various components of the network, while $\boldsymbol{b}_i^{zx}$, $\boldsymbol{b}_i^{yx}$, $\boldsymbol{b}_i^{gx}$ and $\boldsymbol{b}_i$ represent the corresponding bias vectors. The activation functions for layer $i$ are given by $a_i^{x}(\cdot)$, $a_i(\cdot)$ and $a_i^{yx}(\cdot)$. 
Inputs to the network are $\boldsymbol{x}_0 = \mathbf{z}$ and $\boldsymbol{y}_0 = \mathbf{g}$, and the initial weight $\widetilde{\boldsymbol{W}}_0^y$ is set to $\mathbf{0}$. The symbol $\circ$ represents the element-wise product (Hadamard product). 

The integrability of PICINNs requires that the derivatives of $a_i(\cdot), a_i^x(\cdot)$, and $a_i^{y x}(\cdot)$ are common activation functions. Additionally, the convexity of PICINNs with respect to $\mathbf{g}$ is guaranteed by constraining the weights $\widetilde{\boldsymbol{W}}_i^y$ to be non-negative and the activation function $a_i(\cdot)$ to be convex and non-decreasing. Meanwhile, the activation function $a_i^{yx}(\cdot)$ must be non-negative to further uphold convexity requirements. The SoftPlus function satisfies all these requirements, making it the ideal choice for activation functions in PICINNs and is used for EVODMs. To constraint the weights, the function applied to ensure that $W$ is the non-negative transformation of $\widetilde{W}$ is defined as
\begin{equation}
    \widetilde{W} = 
    \begin{cases}
        W + \exp(-\epsilon),
        &W \geq 0, \\
        \exp(W-\epsilon),
        &W < 0.
    \end{cases}
\end{equation}
Here, $\epsilon$ is a positive constant, and it is set to $\epsilon=5$ based on the comparative case study in \cite{sivaprasad2021curious}. This transformation allows $\widetilde{W}$ to take on any real value of the original weight $W$. The resulting $\widetilde{W}$ is then treated as the trainable parameters for the neural networks as shown in Equation~\eqref{eq: picinn}.

\section{Comparisons with VONNs}
\label{appendix: compare vonns}
\begin{figure}[H]
    \centering
    \includegraphics[width=0.8\textwidth]{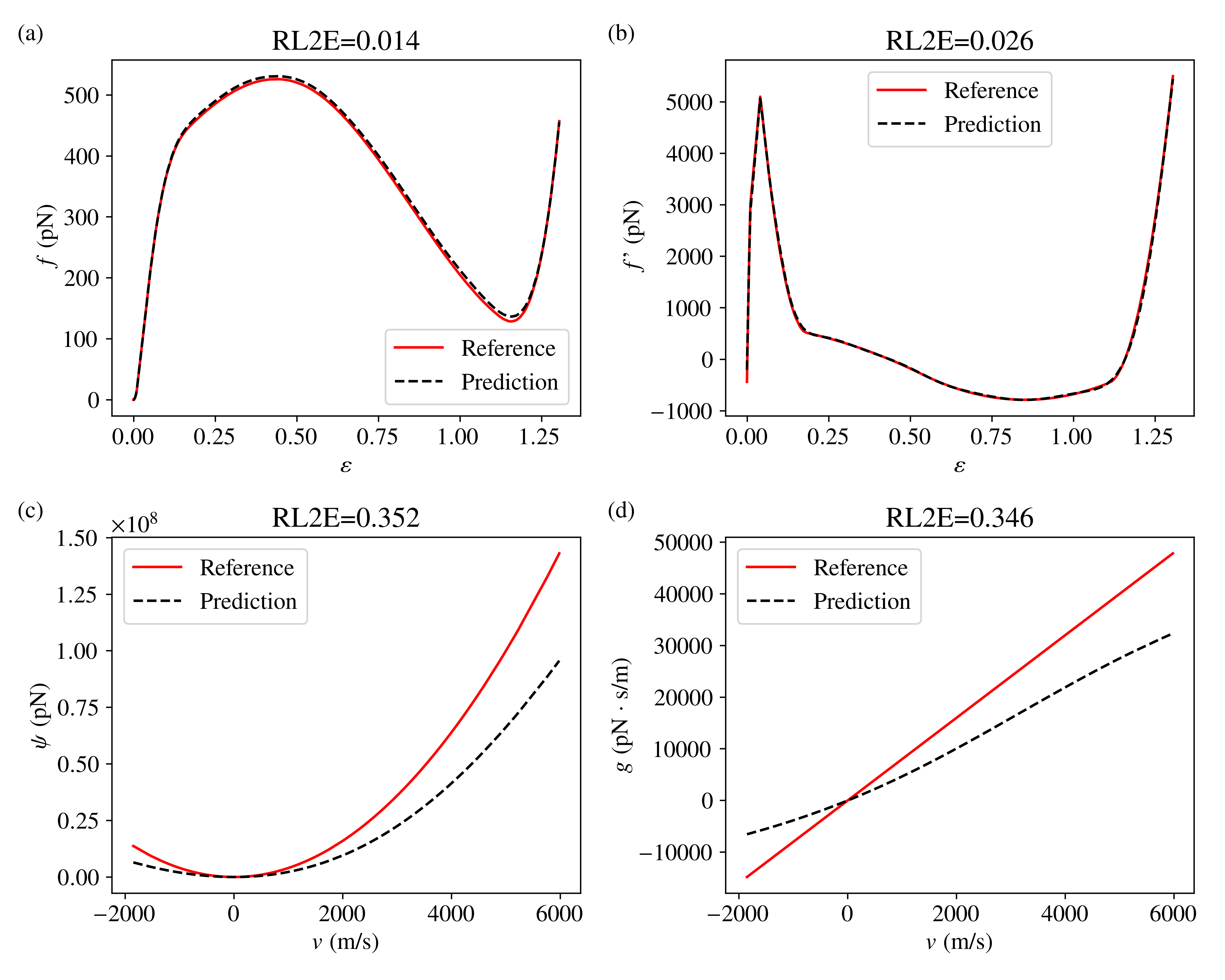}
    \caption{Results obtained from VONNs for the example discussed in Section~\ref{sec: phase transformation}. Comparisons between the reference values and the predictions from VONNs for (a) the free energy density $f(\varepsilon)$, (b) the stress $f'(\varepsilon)$, (c) the dissipation potential density $\psi(v)$, and (d) the viscous force $\psi'(v)$.}
    \label{fig: phase_trans_vonns_results}
\end{figure}

\begin{figure}[H]
    \centering
    \includegraphics[width=0.8\textwidth]{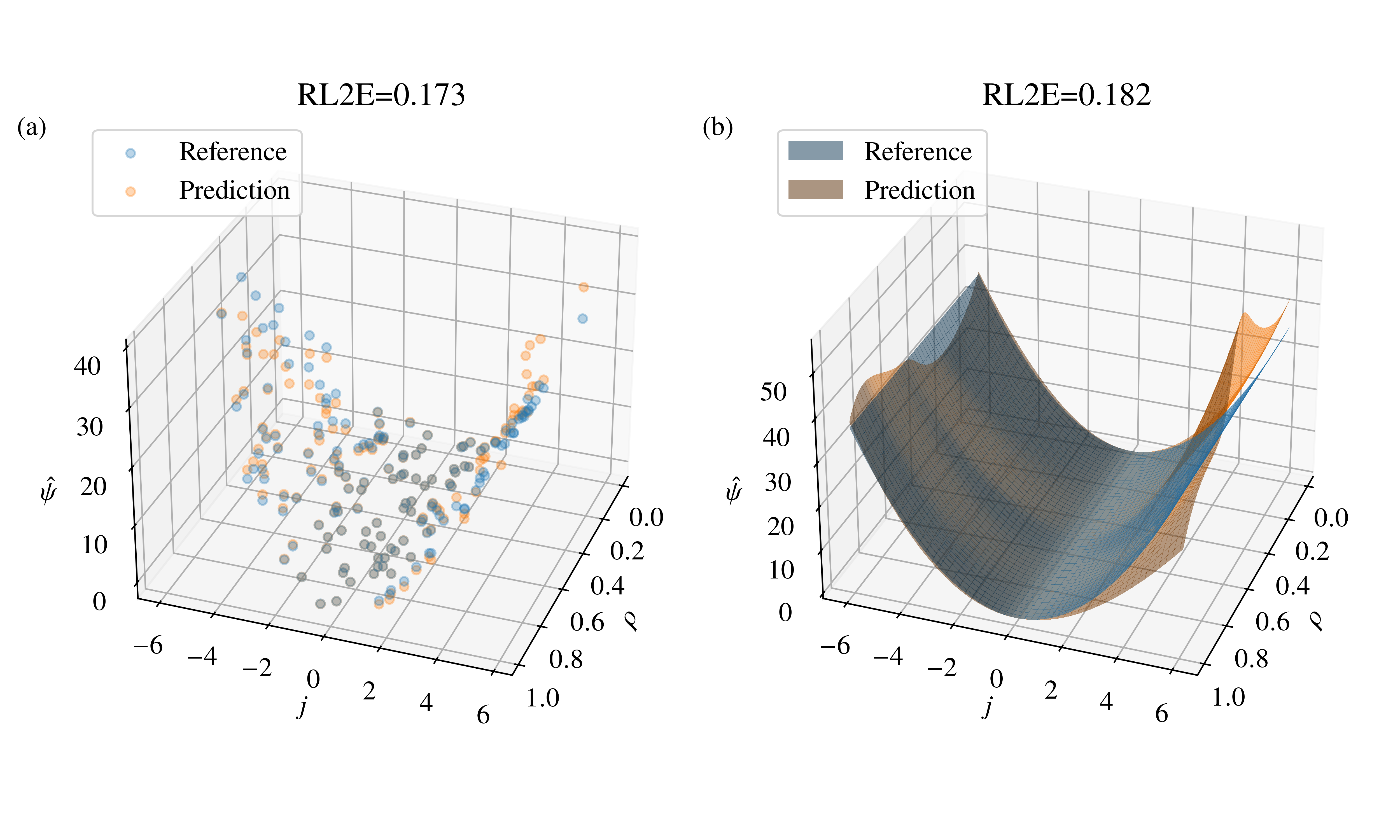}
    \caption{Results obtained from VONNs for the example discussed in Section~\ref{sec: sep}. (a) Comparison between the reference values and the predictions on the testing dataset for $\hat{\psi}$, and (b) comparison between the reference values and the predictions on the full surface for $\hat{\psi}$.}
    \label{fig: sep_vonns_results}
\end{figure}

We train the original VONNs with the same training data used in Section~\ref{sec: phase transformation} and in Section~\ref{sec: sep}. As shown in Figure~\ref{fig: phase_trans_vonns_results} and Figure~\ref{fig: sep_vonns_results}, VONNs fail to achieve competitive accuracy in both examples compared to the proposed EVODMs. Additionally, VONNs are inherently unable to quantify epistemic uncertainty within a single training run.

\section{Training and inference details}
\label{appendix: train details}
All models listed in the following tables are trained using Adam optimizer \cite{kingma2014adam} and Softplus as activation functions. Note that it is important to scale the target data so that DDPMs can distinguish the noise from the data and the noise added in the forward process. Otherwise, the model could be trapped in a local minimum and fail to produce accurate predictions. 
\begin{table}[H]
\centering
\caption{Training details for the base networks of EVODMs and VONNs.}
\begin{tabular}{|c|c|c|c|c|}
\hline
 & \multicolumn{2}{c|}{\textbf{Example 1}} & \multicolumn{2}{c|}{\textbf{Example 2}} \\
\hline
 & \textbf{EVODMs (base)} & \textbf{VONNs} & \textbf{EVODMs (base)} & \textbf{VONNs} \\
\hline
\textbf{Epochs} & 40000* & 40000 & 20000 & 20000\\
\hline
\textbf{Hidden layers ($NN_f$)} & 2 & 2 & 2& 2\\
\hline
\textbf{Hidden layers ($NN_\phi$/$NN_\psi$)} & 2 & 2 & 2& 2\\
\hline
\textbf{Layer size ($NN_f$)} & 50 & 50 & 30 & 30 \\
\hline
\textbf{Layer size ($NN_\phi$/$NN_\psi$)} & 30 & 30 & 30 & 30 \\
\hline
\textbf{Learning rate} & $1 \times 10^{-4}$ & $1 \times 10^{-4}$ & $1 \times 10^{-4}$ & $1 \times 10^{-4}$ \\
\hline
\textbf{Time embedding} & Fourier & & Fourier & \\
\hline
\textbf{Time embed. size} & 2 & & 2 & \\
\hline
\textbf{No. of timestep} & 100 & & 50 & \\
\hline
\textbf{$\beta$ schedule} & Cosine \cite{nichol2021improved} & & Cosine \cite{nichol2021improved} & \\
\hline
\end{tabular}
\label{tab: base_hyperparameters}

\smallskip
\small{*During the first $30000$ epochs, only the BC loss term is used to update the network parameters, in accordance with the two-phase training protocol described in Section~\ref{sec: phase trans nn training}.}
\end{table}
In the inference of the base network, the reverse sampling is sequential and follows $s=S, S-1, ..., 1$. The number of timesteps used in inference is consistent with the number of timesteps used in training.
\begin{table}[H]
\centering
\caption{Training details for Epinets of EVODMs.}
\begin{tabular}{|c|c|c|}
\hline
 & \textbf{Example 1} & \textbf{Example 2} \\
\hline
\textbf{Epochs} & 4000& 5000\\
\hline
\textbf{Hidden layers ($\sigma_f^P$)} & 2 & 2 \\
\hline
\textbf{Hidden layers ($\sigma_\phi^P$)} & 2 & 2 \\
\hline
\textbf{Hidden layers ($\sigma_f^L$)} & 2 & 2 \\
\hline
\textbf{Hidden layers ($\sigma_\phi^L$)} & 2 & 2 \\
\hline
\textbf{Layer size ($\sigma_f^P$)} & 2 & 5 \\
\hline
\textbf{Layer size ($\sigma_\phi^P$)} & 5 & 5 \\
\hline
\textbf{Layer size ($\sigma_f^L$)} & 5 & 15 \\
\hline
\textbf{Layer size ($\sigma_\phi^L$)} & 5 & 15 \\
\hline
\textbf{Learning rate} & $5 \times 10^{-4}$& $5 \times 10^{-4}$ \\
\hline
\textbf{Prior scale ($\kappa$)} & 0.3 & 0.3 \\
\hline
\textbf{Epistemic index size ($\boldsymbol{\gamma}_f$)} & 2 & 5 \\
\hline
\textbf{Epistemic index size ($\boldsymbol{\gamma}_\phi$)} & 2 & 5 \\
\hline
\end{tabular}
\label{tab: epinet_hyperparameters}
\end{table}
Although thermodynamic consistency is strongly embedded within the architecture of the base network, it is important to note that such consistency is not strictly enforced in the Epinets for quantifying epistemic uncertainty. Therefore, if strict thermodynamic consistency is a primary requirement, only the base network of the EVODMs should be used.

Finally, we note that the uncertainty bounds quantified by the Epinets could be sensitive to the choice of the hyperparameters in the examples considered in Section~\ref{sec: phase transformation} and Section~\ref{sec: sep}. 

\section*{Code availability}
The code will be made available upon publication.

\section*{Declaration of competing interest}
The authors declare that they have no known competing financial interests or personal relationships that could have appeared to influence the work reported in this paper.

\section*{Acknowledgments}
The authors acknowledge support from the US Department of the Army W911NF2310230, and thank Weilun Qiu for bringing the Epinet paper to our attention.

\section*{Declaration of generative AI and AI-assisted technologies in the writing process}
During the preparation of this work, the author(s) used ChatGPT in order to check the grammar and improve the clarity of sentences. After using this tool/service, the author(s) reviewed and edited the content as needed and take(s) full responsibility for the content of the published article.

\printbibliography

\end{document}